\documentclass[twocolumn,tighten]{aastex63}

\usepackage{booktabs} 
\usepackage{ wasysym } 
\usepackage{enumitem}
\usepackage{graphicx}
\usepackage[caption=false]{subfig} 
\usepackage[modulo]{lineno} 
\submitjournal{ApJ}

\shorttitle{Family Tree of NLSy1 Transients} 
\shortauthors{Frederick et al.}
\graphicspath{{./}{figures/}}

\begin{document}

\title{A Family Tree of Optical Transients from Narrow-Line Seyfert 1 Galaxies}

\correspondingauthor{Sara Frederick}
\email{sfrederick@astro.umd.edu}

\author[0000-0001-9676-730X]{Sara Frederick}
\affiliation{Department of Astronomy, University of Maryland, College Park, MD 20742, USA}

\author[0000-0003-3703-5154]{Suvi Gezari}
\affiliation{Department of Astronomy, University of Maryland, College Park, MD  20742, USA}
\affiliation{Joint Space-Science Institute, University of Maryland, College Park, MD 20742, USA}
\affiliation{Space Telescope Science Institute, Baltimore, MD 21218, USA}

\author[0000-0002-3168-0139]{Matthew J. Graham}
\affiliation{Division of Physics, Mathematics, and Astronomy, California Institute of Technology, Pasadena, CA 91125, USA}

\author[0000-0003-1546-6615]{Jesper Sollerman}
\affiliation{The Oskar Klein Centre \& Department of Astronomy, Stockholm University, AlbaNova, SE-106 91 Stockholm, Sweden}

\author[0000-0002-3859-8074]{Sjoert van Velzen}
\affiliation{Leiden Observatory, Leiden University, P.O. Box 9513,. 2300 RA Leiden, The Netherlands}

\author{Daniel A. Perley}
\affiliation{Astrophysics Research Institute, Liverpool John Moores University, 146 Brownlow Hill, Liverpool L3 5RF, UK
}

\author[0000-0003-2686-9241]{Daniel Stern}
\affiliation{Jet Propulsion Laboratory, California Institute of Technology, 4800 Oak Grove Drive, Mail Stop 169-221, Pasadena, CA 91109, USA}

\author{Charlotte Ward}
\affiliation{Department of Astronomy, University of Maryland, College Park, MD 20742, USA}

\author[0000-0002-5698-8703]{Erica Hammerstein}
\affiliation{Department of Astronomy, University of Maryland, College Park, MD  20742, USA}



\author[0000-0002-9878-7889]{Tiara Hung}
\affiliation{Department of Astronomy and Astrophysics, University of California, Santa Cruz, CA 95064, USA}








\author[0000-0003-1710-9339]{Lin Yan}
\affiliation{Caltech Optical Observatories, California Institute of Technology, Pasadena, CA 91125, USA}





\author[0000-0003-3768-7515]{Igor Andreoni}
\affil{Division of Physics, Mathematics and Astronomy, California Institute of Technology, Pasadena, CA 91125, USA}

\author[0000-0001-8018-5348]{Eric C. Bellm}
\affiliation{DIRAC Institute, Department of Astronomy, University of Washington, 3910 15th Avenue NE, Seattle, WA 98195, USA} 


\author[0000-0001-5060-8733]{Dmitry A. Duev} 
\affiliation{Division of Physics, Mathematics, and Astronomy, California Institute of Technology, Pasadena, CA 91125, USA}





\author{Marek Kowalski}
 \affiliation{Deutsches Elektronen Synchrotron DESY, Platanenallee 6, 15738 Zeuthen, Germany}
\affiliation{Institut f{\"u}r Physik, Humboldt-Universit{\"a}t zu Berlin, D-12489 Berlin, Germany}
\affiliation{Columbia Astrophysics Laboratory, Columbia University in the City of New York, 550 W 120th St., New York, NY 10027, USA}



\author[0000-0003-2242-0244]{Ashish~A.~Mahabal}
\affiliation{Division of Physics, Mathematics and Astronomy, California Institute of Technology, Pasadena, CA 91125, USA}
\affiliation{Center for Data Driven Discovery, California Institute of Technology, Pasadena, CA 91125, USA}
             
\author{Frank J. Masci}
\affiliation{IPAC, California Institute of Technology, 1200 E. California Blvd, Pasadena, CA 91125, USA}

\author[0000-0002-7226-0659]{Michael Medford}
\affiliation{University of California, Berkeley, Department of Astronomy, Berkeley, CA 94720, USA}
\affiliation{Lawrence Berkeley National Laboratory, 1 Cyclotron Rd., Berkeley, CA 94720, USA}







\author{Ben Rusholme}
\affiliation{IPAC, California Institute of Technology, 1200 E. California Blvd, Pasadena, CA 91125, USA}


\author{Richard Walters}
\affiliation{Caltech Optical Observatories, California Institute of Technology, Pasadena, CA 91125, USA}











\newcommand{\renly}{ZTF18abjjkeo}

\newcommand{\ciri}{ZTF19aaiqmgl}

\newcommand{\tyrion}{ZTF18aajupnt}

\newcommand{\stannis}{ZTF19abvgxrq}

\newcommand{\selyse}{ZTF19aailpwl}

\newcommand{\tywin}{ZTF19aatubsj}

\newcommand{\css}{CSS100217}

\newcommand{\dtm}{PS16dtm}

\newcommand{\bgt}{AT2017bgt}


\newcommand{\wevers}{AT2018fyk}

\newcommand{\jd}{ASASSN-18jd} 

\newcommand{\clagn}{changing-look AGN}

\newcommand{\num}{five}

\begin{abstract}
The Zwicky Transient Facility (ZTF) has discovered five new events belonging to an emerging class of AGN undergoing smooth flares with large amplitudes and rapid rise times. This sample consists of several transients that were initially classified as supernovae with narrow spectral lines. However, upon closer inspection,  all of the host galaxies display resolved Balmer lines characteristic of a narrow-line Seyfert 1  (NLSy1) galaxy. The transient events are long-lived, over 400 days on average. We report UV and X-ray follow-up of the flares and observe persistent UV-bright emission, with two of the five transients detected with luminous X-ray emission, ruling out a supernova interpretation. We compare the properties of this sample to previously reported flaring NLSy1 galaxies, and find that they fall into three spectroscopic categories: Transients with 1) Balmer line profiles and Fe II complexes typical of NLSy1s, 2) strong He II profiles, and 3) He II profiles including Bowen fluorescence features. The latter are members of the growing class of AGN flares attributed to enhanced accretion reported by Trakhtenbrot et al. (2019). We consider physical interpretations in the context of related transients from the literature. For example, two of the sources show high amplitude rebrightening in the optical, ruling out a simple tidal disruption event scenario for those transients. We conclude that three of the sample belong to the Trakhtenbrot et al. (2019) class, and two are TDEs in NLSy1s. We also aim to understand why NLSy1s are preferentially the sites of such rapid enhanced flaring activity.

\end{abstract}



\section{Introduction} \label{sec:intro}
A galaxy center hosting an active galactic nucleus (AGN) is dominated by its continuum emission. 
Therefore, a flare originating from this nuclear region requires a distinctly powerful event to be detectable above this stochastically variable continuum. 
A small number of rapid\footnote{We refer to flare timescales as ``rapid'' when they occur on week to month timescales.}, smoothly evolving flares have been observed to be associated with AGN (e.g. \citealt{Drake2011,Blanchard2017}), with few known mechanisms that can cause these events to occur. 

Intrinsic UV/optical flares, such as those due to enhanced accretion onto the central supermassive black hole (SMBH) in the form of gaseous material or stars passing too close to the nucleus, have been observed in the form of: tidal disruption events (e.g. \citealt{Gezari2012,vanVelzen2020b}), UV-bright flaring events that are associated with accretion rate changes \citep{Trakhtenbrot2019a}, transients with double peaked line profiles linked to accretion disk emission (e.g. Halpern \& Eracleous 1994), or changing-look AGN --- the dramatic change in spectroscopic AGN classification following a rise in continuum level, thought to be connected to unstable changes in accretion state (e.g. \citealt{Lamassa2014,Runnoe2016,MacLeod2016,Ruan2016,Stern2018,Ross2018,Trakhtenbrot2019b,Frederick2019,Graham2020}). 

Phenomena extrinsic to the SMBH accretion engine, such as microlensing of a quasar by a foreground Galactic source (e.g. \citealt{Lawrence2012}) or slowly evolving  super-luminous supernova (SLSN) explosions,  have also been observed to cause smooth large-amplitude flares from galaxies with AGN \citep{Graham2017}. In rare cases these can be astrometrically indistinguishable from the galactic nucleus, and therefore it becomes difficult to discern whether an explosive disruption to the accretion flow has occurred, and to differentiate this from AGN variability \citep{Terlevich1992}. 

Multiwavelength approaches are required to disentangle this diverse family of observed flaring behaviors from AGN. In the golden era of time domain astronomy, even with many multichromatic instruments trained on the sky, a number of newly-discovered objects continue to defy placement into a clear-cut observational category.

In order of discovery, we present a photometric class comprised of \num~rapid flares with similar smooth light curve shapes occurring in a subclass of AGN observed by the Zwicky Transient Facility (ZTF) survey: 
\begin{enumerate}[label=\alph*)]
    \item ZTF19aailpwl/AT2019brs ($z = 0.37362$)
    \item ZTF19abvgxrq/AT2019pev ($z = 0.097$) 
    \item ZTF19aatubsj/AT2019fdr ($z = 0.2666$)
    \item ZTF19aaiqmgl/AT2019avd ($z = 0.0296$)
    \item ZTF18abjjkeo/AT2020hle ($z = 0.103$)
\end{enumerate}

In Section~\ref{sec:obs} we present the follow up of these flares. In Section~\ref{sec:analysis} we compare the results of their respective multiwavelength follow up campaigns to observations of a variety of related objects found in recent years, and in Section~\ref{sec:disc} we attempt to place them into a classification scheme based on observational properties, summarized in Section~\ref{sec:concl}. All transients in the sample are referred to by their ZTF alert names throughout. All magnitudes are reported in the AB system and light curves are shown in the observed frame unless otherwise stated. We have adopted the following cosmology: $H_0$ = 70 km s$^{-1}$ Mpc$^{-1}$,  $\Omega_\Lambda$= 0.73 and $\Omega_M$ = 0.27.

\section{Observations}
\label{sec:obs}

The Zwicky Transient Facility Survey \citep{Bellm2019a,Graham2019} is comprised of the automated Palomar 48-inch Samuel Oschin Telescope (P48) as well as the Palomar 60-inch SED Machine (P60 SEDM; \citealt{Blagorodnova2018,Rigault2019}), and has surveyed the Northern Sky with $g$- and $r$-band filters with a 3-night cadence since 2018 \citep{Bellm2019b}. At least 15 images  meeting good quality criteria were stacked to build a coadded reference image of each observing field and quadrant in each filter band. Science images are subtracted by their references and processed each night by the Infrared Processing and Analysis Center (IPAC) pipeline \citep{Masci2019}. The candidate transient alert stream \citep{Patterson2019} is distributed by the University of Washington Kafka system, and filtered through the AGN and black holes Science Working Group's Nuclear Transients\footnote{A nuclear transient was defined as that within 0.5" of the reference galaxy center. Over 9000 nuclear transients passed this filter and were ranked during ZTF Phase I, of which 27 were TDEs, over 7\% were classified as SN, and over half were AGN or candidate AGN.} parameter criteria (outlined in \citealt{vanVelzen2019,vanVelzen2020}) by the Ampel broker \citep{Nordin2019,Soumagnac2018}, with the GROWTH Marshal user interface utilized for the coordination of follow-up efforts \citep{Kasliwal2019}. 

All 5 transients included in the sample presented here were selected based on the following criteria: large amplitude, nuclear variability ($\Delta g > 1$ mag in difference imaging photometry, and within 0.5\arcsec~of the center of the host galaxy in the reference image) with follow-up or pre-flare spectra consistent with an AGN classification. This selection was not systematic (and therefore not complete), but rather the result of ongoing intersecting and collaborative searches for changing look AGN \citep{Frederick2019}, TDEs \citep{vanVelzen2019,vanVelzen2020}, and superluminous supernovae \citep{Lunnan2019,Yan2020} relying on partial human vetting from the ZTF transient alert stream, from which this sample emerged as more examples were collected. A systematic search for NLSy1 transients in ZTF will be the focus of a future study. 

\subsection{Optical Photometry} 

\begin{figure*}[ht!]
\includegraphics[scale=0.95]{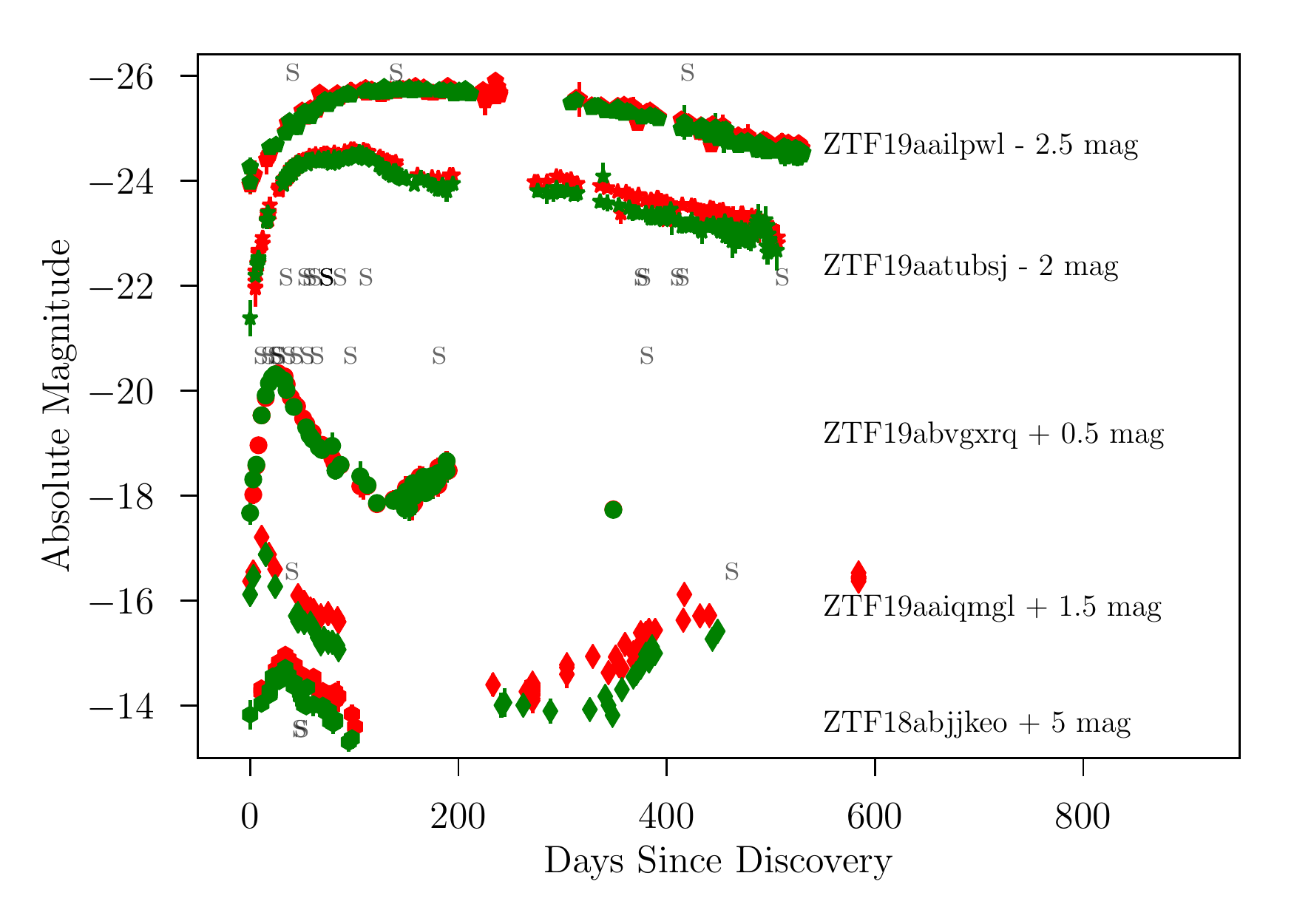}
\caption{Comparison of the ZTF $g$- and $r$-band difference imaging light curve shapes and absolute magnitudes of the sample.  \tywin~decreases before reaching a second plateau stage, and undergoes significant reddening after the first plateau while the others never do.  \stannis~rises again symmetrically after decreasing to pre-flare levels, as does \ciri. The light curves have been shifted in absolute magnitude space for visual purposes, as indicated alongside the object name. Overlap of the $g$ and $r$ light curves reflects true colors such that the initial colors approach $g-r=0$ mag for all transients in the sample. Observations at other wavelengths are shown in Figure~\ref{fig:nulnulc}. Spectroscopic epochs are labeled for each light curve with an `S' below \tywin~and \renly~and above the rest. \\
}
\label{fig:lc}
\end{figure*}

All transients in the sample were detected pre-peak using ZTF difference imaging photometry. 
The smooth light curve shapes (with scatter $\Delta g < 0.1$ mag) of the sample are shown in Figure~\ref{fig:lc}. All magnitude changes are reported in $g$ band unless otherwise noted. An analysis of the rise times to peak are measured and reported in Section~\ref{sec:timescales}. We report the $g$-band magnitude-weighted offsets for each transient, calculated using Equation 3 in \citet{vanVelzen2019}.  
ZTF forced photometry for the sample is shown in Figure~\ref{fig:ipac_lc} of the Appendix.

{\it\selyse} --- (RA=14:27:46.41, Dec=+29:30:38.6, J2000.0) was first detected on 2019 Feb 08 as a nuclear transient within 0\farcs17 of the host galaxy center. 
The host galaxy displayed some variability at the $<1$ mag level in the Catalina Real-Time Transient Survey (CRTS; \citealt{Drake2009}) from 2005 to 2013.

{\it\stannis}\footnote{\stannis~passed the ZTF TDE working group's tidal disruption event criteria, and was given the nickname ``Stannis Baratheon" for ease of discussion. When it was found to be among a class of AGN-associated objects serendipitously detected by ZTF, the other sources in the class were retroactively given the names of other Game of Thrones characters in the same Great House - and collectively referred to fondly as ``The Baratheons'', whose motto is, appropriately, ``Ours is the Fury''. } --- (RA=04:29:22.72, Dec=+00:37:07.6, J2000.0), also known as Gaia19eby, was first detected  on 2019 Sept 04 as a nuclear transient within 0\farcs15 of the host galaxy center. 
ATLAS, Gaia, and PanSTARRs also reported observations of this source on the Transient Name Server (TNS) with discovery dates of 2019 Sep 04, 2019 Sep 13, and 2019 Sep 26, respectively. 
The host galaxy displayed no variability above the 0.5 mag level in CRTS. 

{\it\tywin} --- (RA= 	17:09:06.86, Dec=+26:51:20.7, J2000.0) was detected on 2019 Apr 27 with a significant flux increase with respect to the reference image and with an offset from the nucleus of its host of 0\farcs13. 
During a coverage gap in the first 40 days of the rise, ATLAS reported an intriguing ``bump'' feature \citep{Smartt2019}. 
The host galaxy displayed variability at the 2 mag level in V-band CRTS data from 2009 to 2013 (variability which was not observed in ZTF forced photometry prior to the transient). 

{\it\ciri} --- (RA=08:23:36.77, Dec=+04:23:02.5, J2000.0), also known as eRASSt J082337+042303\footnote{This was the only source in the sample to be detected by the extended ROentgen Survey with an Imaging Telescope Array (eROSITA, part of the Russian-German "Spectrum-Roentgen-Gamma" (SRG) mission; \citealt{Cappelluti2011}), and was given the name eRASSt J082337+042303. This X-ray detection coincident with the transient's host galaxy is described in Section~\ref{sec:obs:xray}.}, was detected by ZTF beginning on 2019 Feb 09 within 0\farcs06 of its host galaxy. 
The host showed no variability in CRTS for 15 years prior to its rapid rise to peak. 

{\it\renly} --- (RA=11:07:42.91, Dec=+74:38:02.0, J2000.0) was detected beginning on 2020 Apr 05 within 0\farcs02 of its host galaxy center. 
The ZTF forced photometry for this source shows no variability above the level of the galaxy for $>400$ days. The host galaxy of \renly~was beyond the survey limits of CRTS.

\subsection{Optical Spectroscopy}

All spectroscopic follow-up observations for the sample are summarized in Table~\ref{tab:spc}, and each epoch is shown in Figure~\ref{fig:heii_spc} of the Appendix. The phases of the optical follow-up spectra with respect to the features in the ZTF light curves are annotated on Figure~\ref{fig:lc}. All transients in this sample have spectral  characteristics of NLSy1 galaxies, i.e. strong Balmer line emission with FWHM $<2000$ km s$^{-1}$, along with other spectral features which are highlighted below and explored in detail in Section~\ref{sec:anal:spc}. 
We reduced Palomar 60" SED Machine (P60/SEDM; Program PIs: Gezari, Sollerman, Kulkarni) spectra with \texttt{pysedm} \citep{Rigault2019}, and all other spectra with \texttt{pyraf} using standard procedures.

\setlength{\tabcolsep}{2pt}
\begin{table*}[ht!]

\caption{Summary of spectroscopic follow-up observations of the sample.
} 
\label{tab:spc}
\begin{tabular}{@{}lllll@{}}
\toprule
Name & Obs UT       & Instrument              & Exposure (s) & Reference   \\ \toprule
\selyse & 2006 Jul 01  & SDSS & 3000         & \citet{Abolfathi2018}                                \\
& 2019 Mar 15 & FLOYDS-N               & 3600        & \citet{Arcavi2019}                               \\

& 2019 Jun 22 & LDT Deveny              & 900        & This work                                \\


\midrule
\tywin & 2019 May 25  & Palomar 60" SEDM & 2250         & This work                                \\
& 2019 Jun 17  & LT SPRAT & 900         & This work                                \\
& 2019 Jun 22 & LDT Deveny              & 900         & This work                                \\
& 2019 Jul 03   & Palomar 200" Hale             & 600          & This work                                \\ 
 & 2020 Apr 30 & NOT ALFOSC & 1750 & This work \\



\midrule
\stannis & 2019 Sep 08  & Palomar 60" SEDM & 2250         & This work                                \\
& 2019 Sep 15  & LT SPRAT & 500         & This work                                \\
& 2019 Sep 22  & Palomar 60" SEDM & 2250         & This work                                \\
& 2019 Sep 24 & LDT Deveny              & 600         & This work                                \\
& 2019 Sep 25   & Keck LRIS             & 300          & This work                                \\ 
& 2019 Sep 25 & NICER & 2000 & \citet{Kara2019ATel} \\
& 2019 Oct 01 & Chandra LETG & 45400 & \citealt{Miller2019} \\
& 2019 Oct 05   & Lick 3-m KAST             & 1500          & This work                                \\ 
& 2019 Oct 12  & LT SPRAT & 500         & This work                                \\
& 2019 Oct 15  & Chandra LETG & 91000 & \citealt{Mathur2019} \\
& 2019 Oct 23 & LDT Deveny              & 900         & This work                                \\
& 2019 Nov 01  & Palomar 60" SEDM & 2250         & This work                                \\
& 2019 Dec 03 & LDT Deveny              & 2400         & This work                                \\
& 2020 Feb 26 & LDT Deveny              & 2600         & This work                                \\
& 2020 Jan 30 & {\it Swift XRT}               & 94700        & This work                                \\

\midrule
\ciri & 2020 Mar 15 & NOT ALFOSC & 1800 & Malyali et al. 2020, in prep. \\ 
& 2020 Apr 28 & eROSITA SRG & 140 & Malyali et al. 2020, in prep. \\ 
& 2020 May 10 & FLOYDS-S & 3600 & \citet{Trakhtenbrot2020} \\

\midrule
\renly & 2020 May 16 & Palomar 60" SEDM & 2250 & This work \\ 
& 2020 May 18 & LT SPRAT & 1000 & This work \\ 
\bottomrule

\end{tabular}
\end{table*}


{\it\selyse} --- showed a striking difference to the SDSS spectrum showing it was a NLSy1 as early as 2006 \citep{Abolfathi2018,Rakshit2017}. The follow-up Folded Low Order whYte-pupil Double-dispersed Spectrograph North (FLOYDS-N;  \citealt{Arcavi2019} and Lowell Discovery Telescope (LDT, formerly DCT; PI: Gezari) spectra showed a steep blue continuum and a strong He~II profile with Bowen fluorescence features, indicating it became a flaring SMBH belonging to the observational class established by \citet{Trakhtenbrot2019a}.

{\it\stannis} --- was spectroscopically identified as a NLSy1 on 2019 Sept 15 with the Liverpool Telescope (LT; PI: Perley) SPectrograph for the Rapid Acquisition of Transients (SPRAT), based on the width of the Balmer emission lines and the strength of the [O~III]~$\lambda$5007 emission line. \citet{Gezari2019} reported that the LT spectrum showed evidence for blue-shifted He~II~$\lambda$4686 emission as well as N~III~$\lambda$4640 emission, due to the Bowen fluorescence mechanism, placing it again in the observational subclass of the \citet{Trakhtenbrot2019a} objects. 
Near peak it was observed with Keck 10-m Low Resolution Imaging Spectrometer (LRIS; PI: Graham) as well as the LDT Deveny Spectrograph (PI: Gezari) and the KAST Double Spectrograph on the Lick 3-m Shane Telescope (PI: Foley), which confirmed the strong blue continuum and clearly defined and persistent Bowen fluorescence features.

{\it\tywin} --- was observed 8 days after peak on 2019 Jul 03 with the Double Spectrograph (DBSP) on the Palomar 200-inch Hale Telescope (P200; PI: Yan). We measured a significant ``blue horn'' component of H$\beta$ and marginally detected He~II. The transient continuum of \tywin~faded to reveal an underlying Fe~II complex in the Nordic Optical Telescope (NOT; PI: Sollerman) spectrum taken nearly 368 days after peak on 2020 Apr 30, with no evidence for He~II emission. 

{\it\ciri} --- The spectrum taken with NOT (PI: Sollerman) on 2019 Mar 15 near the first optical peak showed strong Balmer line emission, no detection of a He~II line complex, and evidence for a Fe~II complex, characteristic of NLSy1 galaxies.  A follow-up FLOYDS-S spectrum taken  444 days after peak and reported to the Transient Name Server (TNS) by \citet{Trakhtenbrot2020} showed the appearance of He~II and Bowen fluorescence features and a ``blue horn'' in H$\beta$. Again this event was classified as a member of the \citet{Trakhtenbrot2019a} observational class of flaring NLSy1s. 

{\it\renly} --- In the LT (PI: Perley) spectrum of \renly~taken on 2020 May 18 8 days after peak, the narrow component of the He~II profile is significantly blueshifted. No Fe~II line complex was detected in the spectra of this transient. 

\subsection{UV Photometry} 

We triggered target-of-opportunity monitoring observations with the Neil Gehrels {\it Swift} Telescope \citep{Gehrels2004} for all transients in the sample. Using the HEASOFT command \texttt{uvotsource} we extracted {\it UVOT} photometry within a 5\arcsec-radius circular aperture and using an annular background region centered on the coordinates of the optical transient. 

Figure~\ref{fig:nulnulc} shows the $\nu L_\nu$ light curves of all flares in the sample. We compare ZTF $g$ and $r$ band difference imaging, WISE difference imaging, {\it Swift XRT} monitoring, and {\it Swift UVOT} detections subtracted by the archival {\it Galaxy Evolution Explorer} ({\it GALEX};  \citealt{Bianchi2017}) All-Sky Imaging Survey (AIS) near-UV (NUV, $\lambda_{\rm{eff}}=2310$ \AA) host measurements (measured with a 6-\arcsec radius aperture). 

We found all transients in the sample to be UV-bright, but with varying UV colors. The UV color of \ciri~($UVW1-g=-0.2$ mag) was similar to that of \stannis~($UVW2-g=-0.2$ mag) and \selyse~(ranging from $UVW2-g=-0.1$ mag to $-0.7$ mag in 80 days), but \tywin~was the only transient in the sample with positive UV color ($UVW2-g=0.8$ mag). 
 The UV light curves of the sample tend to follow the shape of the optical. \stannis~became host dominated over time as the transient faded. but with strong scatter in the light curve as it approached the host magnitude. 



 \subsection{X-rays} \label{sec:obs:xray}
 
We found only two transients in the sample to be X-ray bright in follow-up {\it Swift XRT} observations: \stannis~and \ciri. \selyse~was  detected only once, and then only at a low level. We measured an {\it XRT} upper limit of 0.004 counts s$^{-1}$ for \tywin.
 X-ray follow-up spectra are reported in Table~\ref{tab:spc}. {\it Swift} photometry compared to WISE W1- and W2-band and ZTF $g$ and $r$-band photometry is shown in Figure~\ref{fig:nulnulc}. The X-ray bright flares in this sample tend to vary in lockstep with the slow UV/optical flares.



{\it\selyse} --- was detected only once in 11 observations during a 16-month monitoring campaign between 2019 Mar 21 and 2020 Jul 7.  We measured a 3-$\sigma$ detection of  0.003 counts s$^{-1}$ on 18 Apr 2019, just brighter than the limiting flux.

{\it\stannis} --- Similar to the UV light curve, the shape of the X-ray flare of \stannis~followed the optical, from its fade through its second rise (See Figure~\ref{fig:nulnulc} and Section~\ref{sec:analysis}). The unabsorbed $0.3-10$ keV flux from the stacked {\it XRT} spectrum of \stannis~was $7.3\pm0.1 \times 10^{-12}$ erg cm$^{-2}$ s$^{-1}$. \stannis~was previously detected in ROSAT, and NICER  observations show an increase in flux from this by 100 times  ($1\times10^{-11}$ erg cm$^{-2}$ s$^{-1}$), variable from 11 to 14 counts s$^{-1}$ in 3 hours \citep{Kara2019ATel}. A 50 ks Chandra LETG grating observation taken just 8 days after peak and reported by \citet{Miller2019} found a flux consistent with this, with the spectral shape a good fit to a  $kT=0.24$ keV blackbody, and the source variable at the 25\% level on $2-3$ ks timescales. \citet{Mathur2019} reported a decrease in $0.3-2.5$ keV flux to $7.7\times 10^{-12}$ erg cm$^{-2}$ s$^{-1}$; their 91 ks Chandra LETG observation was a good fit to a consistent blackbody model and a power law component typical of AGN with spectral index $\Gamma=1.8$, with no  intrinsic absorption required. 


{\it\ciri} --- was observed only during the second optical flare (on 2020 Apr 28, 350 days after the first ZTF detection), and was the only X-ray bright transient in the sample with much fainter X-ray $\nu L_\nu$ than that of the optical (shown in Figure~\ref{fig:nulnulc}). Like \stannis, the shape of the X-ray rise followed that of the second rise. It was detected by eROSITA as eRASSt J082337+042303, a soft X-ray transient consistent with the galaxy 2MASX J08233674+0423027 (\citealt{Malyali2020ATel}, Malyali et al. 2020, in prep.) Prior to this, the XMM Slew Survey reported a non-detection at the location of the host galaxy, with an upper limit of $<1.7 \times 10^{-14}$ erg$^{-1}$ s$^{-1}$ cm$^{-2}$ assuming $kT_{\rm{bb}}=100$ eV and $N_H=3 \times 10^{20}$ cm$^{-2}$. The SRG flux of $1.5\times 10^{-12}$ erg$^{-1}$ s$^{-1}$  cm$^{-2}$ was 90 times brighter than this upper limit. No hard X-ray component was detected above 2.3 keV. No strong short-term variability on hours-long timescales was detected, and no strong variability was detected between SRG and the 3 {\it Swift XRT} monitoring observations taken afterward with a week-long cadence. Swift and NICER observations over the next 5 months showed an additional increase in X-ray flux by a factor of 10 \citep{Pasham2020}. A careful study of the X-ray properties of this transient is forthcoming (Malyali et al. 2020, in prep.)

\subsection{IR} 
\citet{Malyali2020ATel} reported that the WISE color of \ciri~was atypically low ($W1-W2 \simeq 0.07$ mag) compared to typical AGN values  ($W1-W2 = 0.7-0.8$ mag, \citealt{Stern2012,Assef2013}). 
The WISE colors of \renly~(neoWISE: 0.35 mag, AllWISE: 0.036 mag) and \stannis~($W1-W2 = 0.45$ mag) are also inconsistent with an AGN, though not quite as low as that of \ciri. Only \selyse~truly appeared as an AGN in IR, with $W1-W2=0.98$ mag. The WISE AGN classification of the sample is summarized in Table~\ref{tab:properties} in Section~\ref{sec:disc}.

\begin{figure*}[ht!]
\includegraphics[scale=0.52]{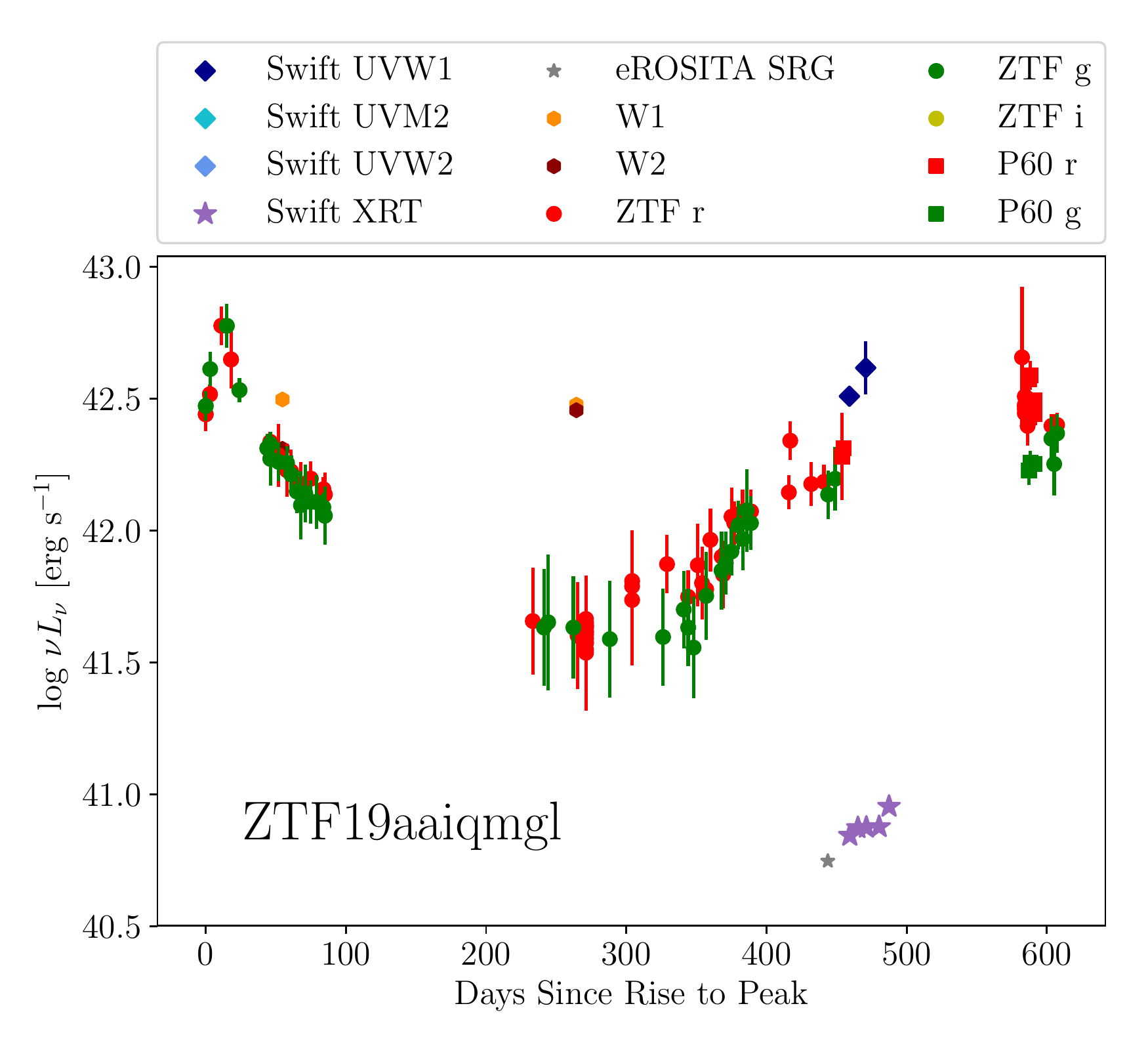}
\includegraphics[scale=0.52]{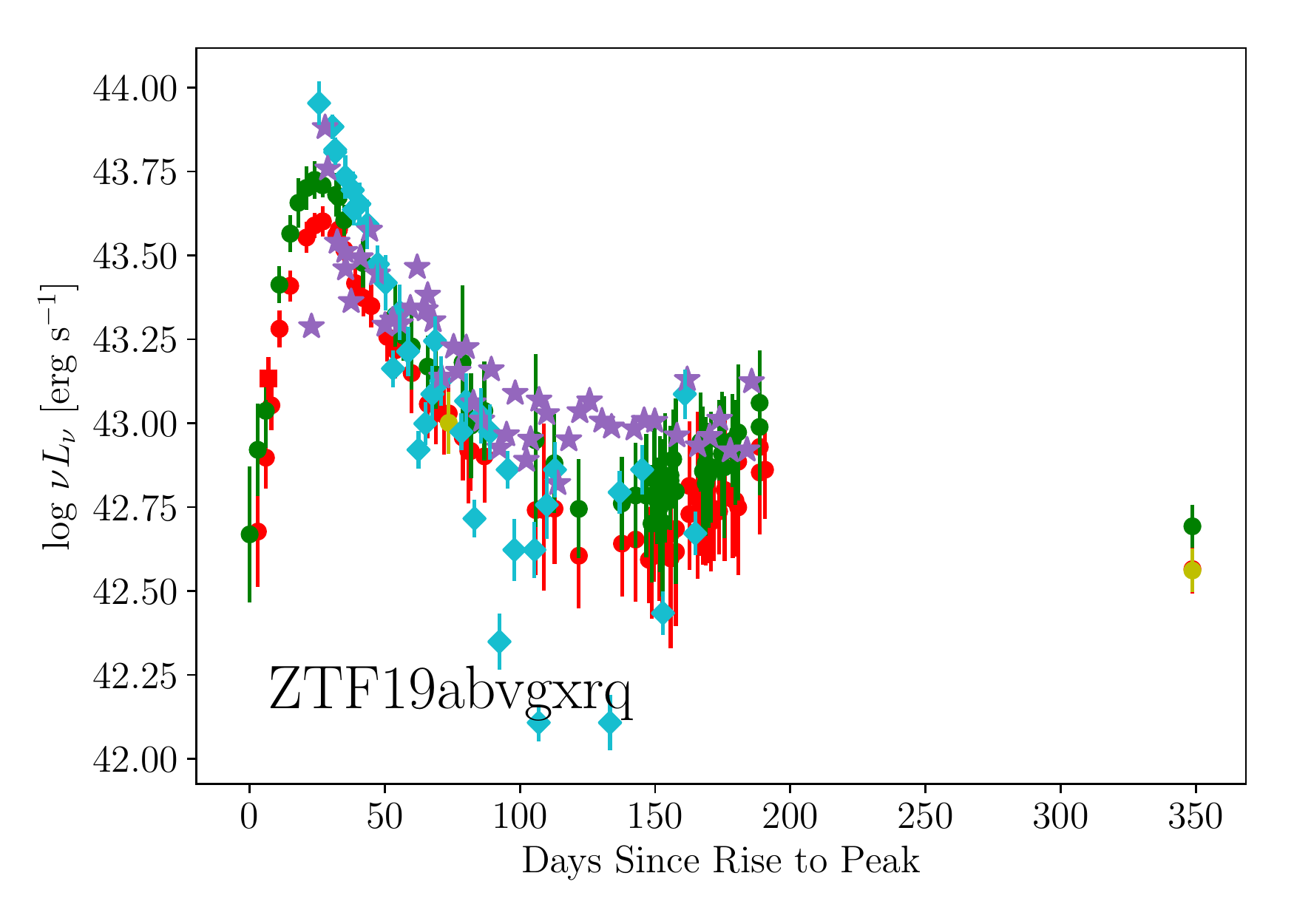}
\includegraphics[scale=0.52]{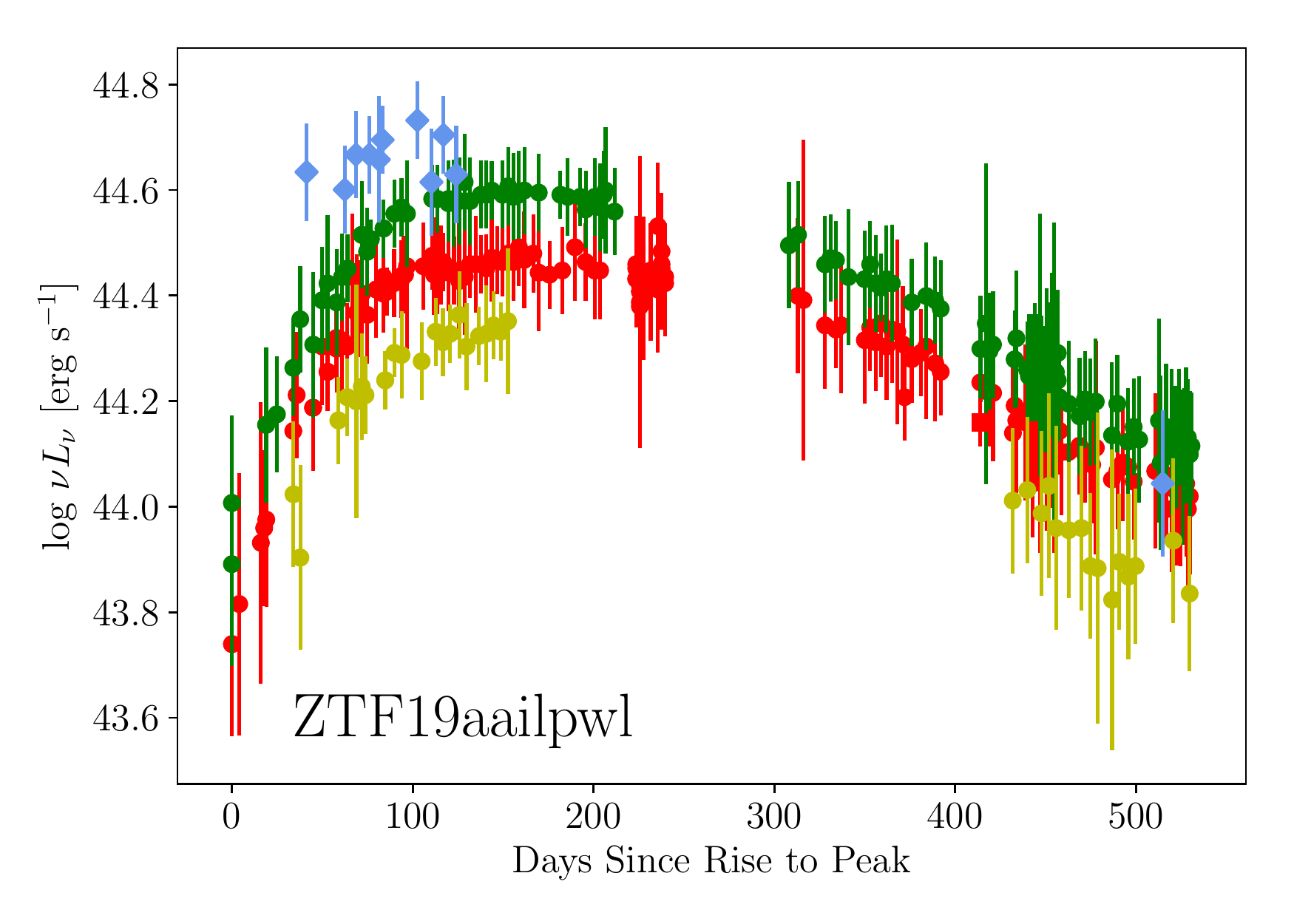}
\includegraphics[scale=0.52]{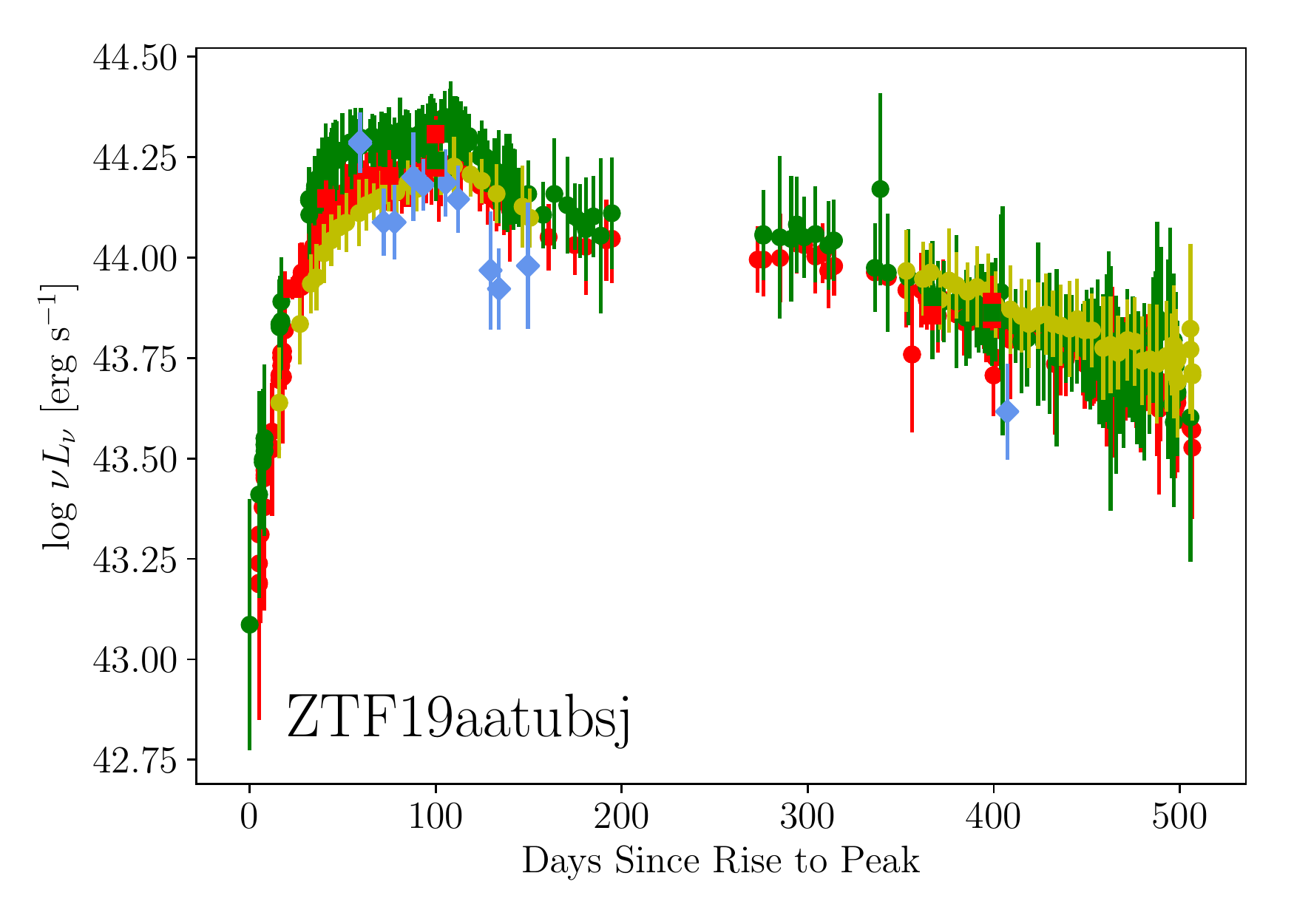}
\includegraphics[scale=0.52]{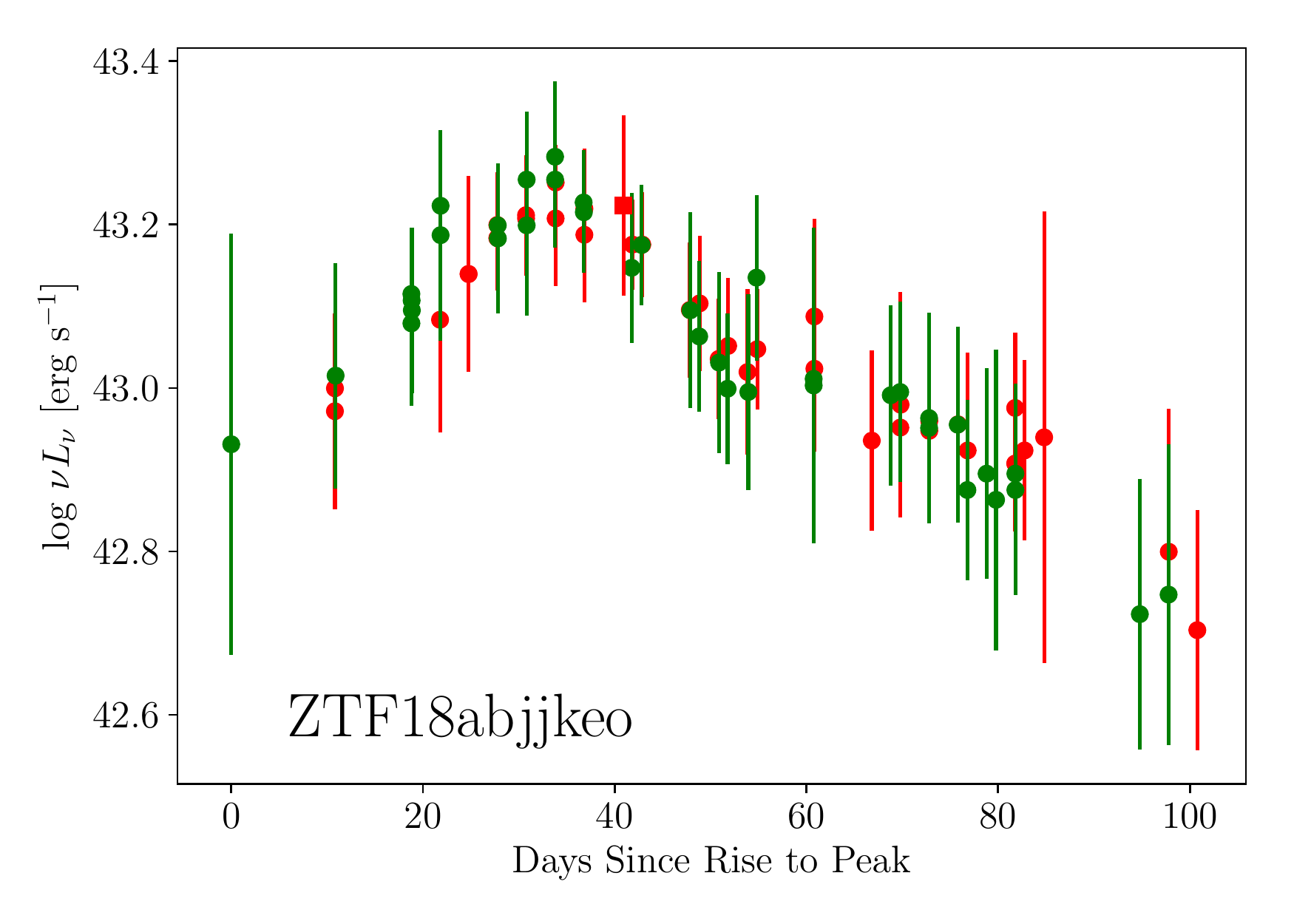}

\caption{We track the colors of the transients in the sample with a $\nu L_\nu$ light curve, comparing the ZTF and WISE data to concurrent high cadence {\it Swift UVOT} and {\it XRT} monitoring observations. The X-ray rise and fade of \stannis~tracks the optical/UV with no significant delay. We subtracted the host galaxy light as estimated by {\it GALEX} NUV measurements from the {\it Swift UVOT} observations. Times are given in days since first ZTF detection. The X-ray errorbars are comparable to the size of the data points. See Figure~\ref{fig:ipac_lc} for pre-outburst forced photometry. } 

\label{fig:nulnulc}
\end{figure*}

A flare in the IR was detected in NeoWISE at the location of \ciri~and concurrent with the optical and X-ray transient. Though the IR flare began much sooner in 2009, Figure~\ref{fig:nulnulc} shows that the peak of the flare was delayed with respect to the first optical peak. Prior to this flare, WISE photometry detected no variability at the location of \ciri~for nearly 5 years.


\section{Analysis}
\label{sec:analysis}

%

\subsection{Photometry}

The difference imaging light curves for the sample are shown in terms of absolute magnitudes in Figure~\ref{fig:lc_lit}. 

We show the sample alongside various NLSy1-related events from the literature, which are described in more detail in Section~\ref{sec:disc}.  
\css~displayed some variability prior to the transient, unlike any of the events in this sample. 
\bgt~was observed only during its fade in difference imaging, so we instead show its aperture photometry (from the ASAS-SN Photometry Database\footnote{\url{https://asas-sn.osu.edu/photometry}}; \citealt{Jayasinghe2019}) which also shows the rise of the source. 
\tyrion~is by far the least luminous transient shown. 
We note the similarity of the shapes of the light curves of \tywin~and PS16dtm, which is discussed further in Section~\ref{sec:disc}.

\subsubsection{Light Curve Timescales} \label{sec:timescales}

\begin{figure*}[ht!]
\includegraphics[scale=0.92]{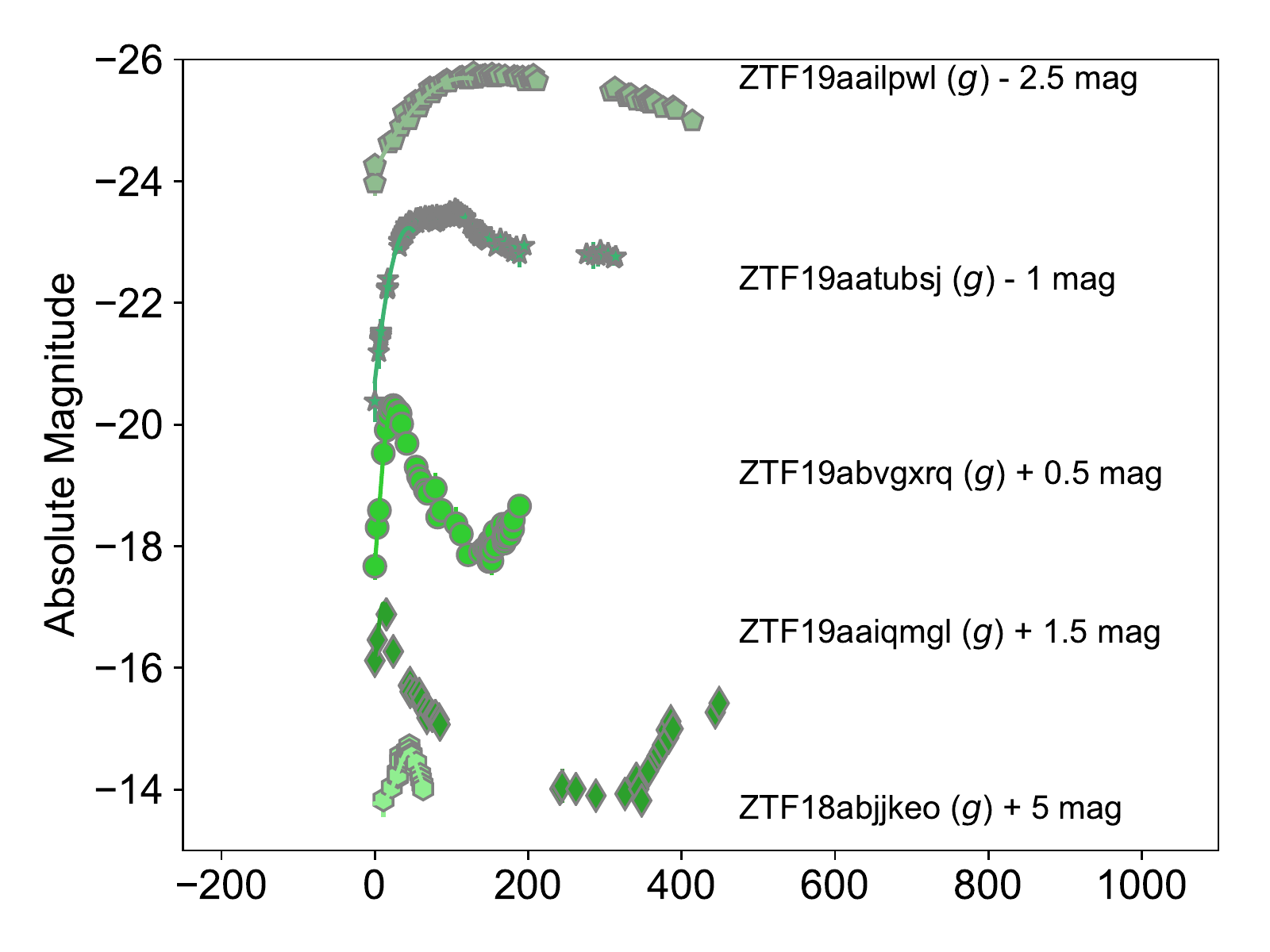}
\includegraphics[scale=.99]{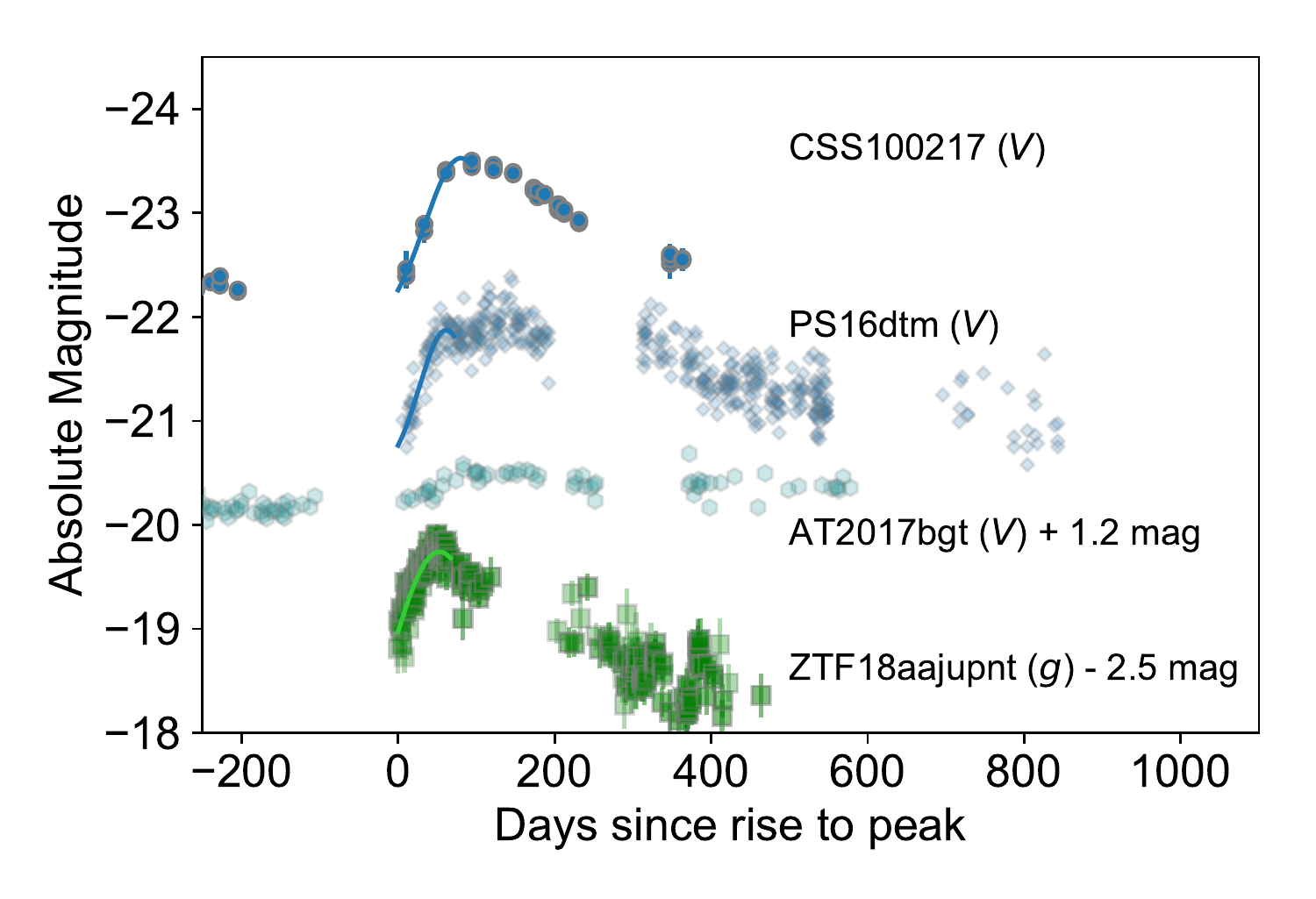}

\caption{The difference imaging light curves of the ZTF sample (upper panel) compared to the published light curves of NLSy1-related events from the literature (lower panel): changing-look LINER \tyrion \citep{Frederick2019}, TDE in a NLSy1 PS16dtm \citep{Blanchard2017}, SN in a NLSy1 CSS100217 \citep{Drake2011}, and the aperture photometry of flaring NLSy1 AT2017bgt \citep{Trakhtenbrot2019a}. We show only $g-$band observations for the ZTF sample (upper panel), and omit errorbars for visual purposes. Note the differences in optical filters shown ($g$ in green, $V$ in blue), the differences in colors and markers used to represent the same filters for visual clarity, as well as the difference in y-axis scale between the panels. CRTS data for \css~$>200$ days prior to the transient is not shown on this scale, but showed no significant activity for $>5$ years.}
\label{fig:lc_lit}
\end{figure*}

\begin{figure*}[ht!]
\centering
\includegraphics[scale=0.65]{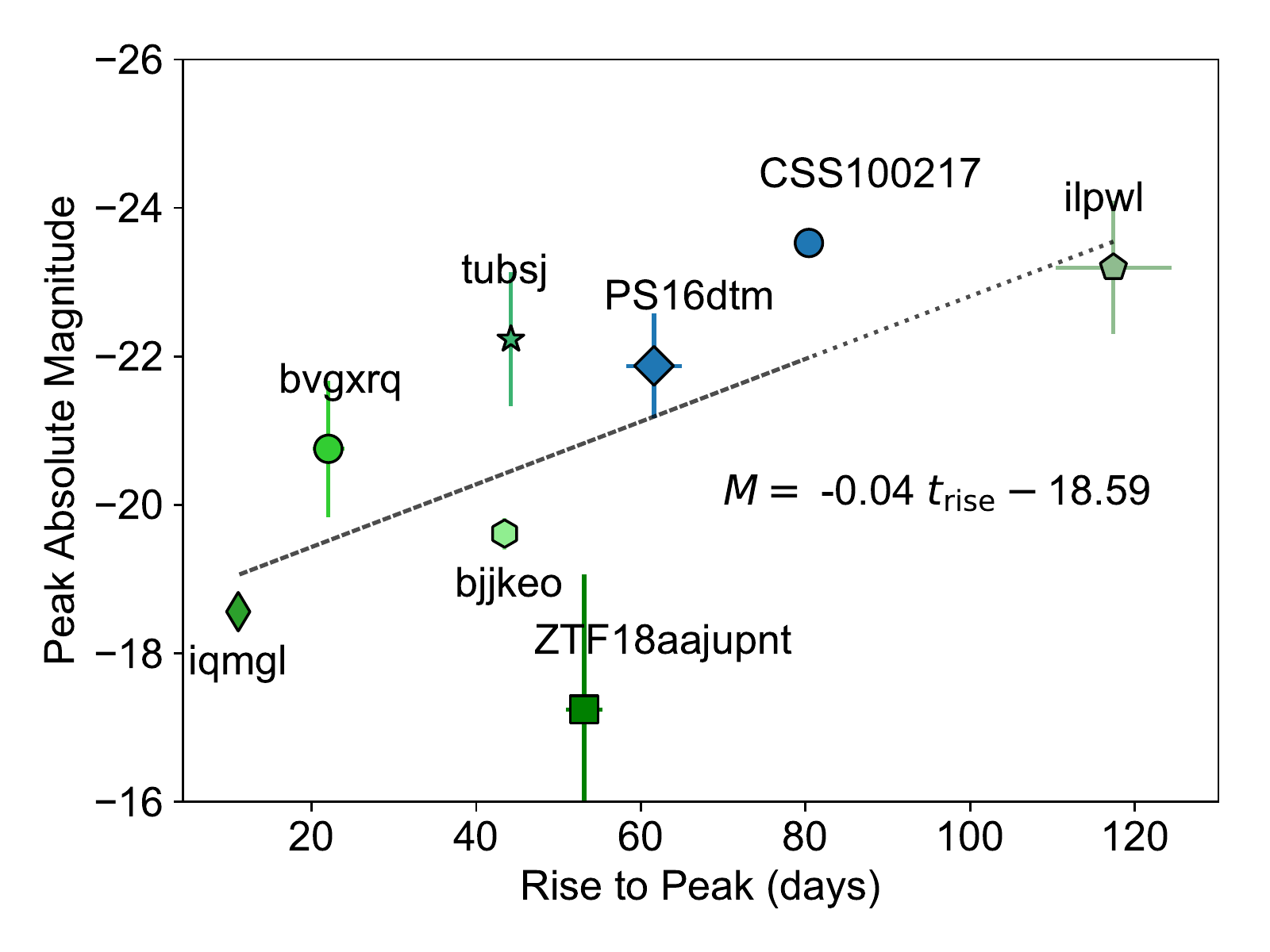}
\caption{Correlation of the rise times of the sample light curves with the maximum absolute magnitude. Fits to light curves are described in Section~\ref{sec:timescales}. The same color scheme and markers are used as in Figure~\ref{fig:lc_lit}.}
\label{fig:rise}
\end{figure*}

We measured the rise-to-peak timescales of the sample by fitting Gaussians to the light curves shown in Figure~\ref{fig:lc_lit} using the \texttt{lmfit} package. 
We observe a correlation between the luminosity (specifically the absolute magnitudes $M_V$ and $M_g$) and rise-to-peak timescales of the sample ($t_{\rm{rise}}$) with the following relation: $M=-0.04 t_{\rm{rise}}-18.59$, shown in Figure~\ref{fig:rise}. Fitting the light curves with quadratic functions resulted in the same correlation within the error estimates. 
Interestingly, TDEs also show a positive correlation between rise time to peak and luminosity \citep{vanVelzen2020}. 
\tyrion~appears under-luminous for how fast it rises. 
AT2017bgt was observed only during its fading phase in difference imaging, and so was excluded from this portion of the analysis. 


\subsubsection{Rebrightening}

It is noteable that two sources in the sample, \stannis~and \ciri, each have a dramatic rebrightening episode. Following a flare and an approximately $\sim$2 mag fade from peak, both return to nearly half their maximum luminosity before seasonal gaps in visibility. This is in contrast to that of almost all TDEs and SN in the literature (e.g. \citealt{Sollerman2019,Sollerman2020}), although they can show plateaus and ``humps'' (e.g. Hammerstein et al. 2020, in prep.)\footnote{We note that ASASSN-15lh showed a large amplitude ``double-humped'' structure in its UV light curve.} 
We explore possible interpretations of this rebrightening in Section~\ref{sec:disc}.

\subsubsection{UV/Optical to X-ray Ratio}\label{sec:aox}
We derive the simultaneous UV/optical-to-X-ray spectral slope ratio ($\alpha_{\rm{OX}}$) from the {\it Swift UVOT} and {\it XRT} observations of the sample, (as well as upper limits assuming $\Gamma_X=2$, when applicable). We compute unabsorbed X-ray flux densities at 2 keV using the \texttt{PIMMS} v4.10 web tool\footnote{\url{https://cxc.harvard.edu/toolkit/pimms.jsp}}. Following Eq. 4 of \citet{Tananbaum1979} and Eq. 11 of \citet{Grupe2010}, the definition of this ratio is $\alpha_{\text{OX}}= 0.3838~\text{log}(L_{\text{2 keV}}/L_{2500\rm{A}})$. 
Of the transients detected in X-rays, the $\alpha_{\rm{OX}}$ of \stannis~evolves over 150 days between 1.1 and 1.4, and \selyse~is observed in X-rays during only one epoch with $\alpha_{\rm{OX}}=1.7$, equivalent to that of the late time detections of \ciri.
The range of $\alpha_{\rm{OX}}$ measured for the sample is consistent with that of NLSy1s ($0.9 < \alpha_{\rm{OX}} < 1.8$; \citealt{Gallo2006}). 


\subsection{Spectroscopy}\label{sec:anal:spc}
\begin{figure*}[ht!]
  \centering

  \subfloat[H$\alpha$ \stannis~0d]{\label{fig:linefit_stannis_ha}\includegraphics[scale=0.33]{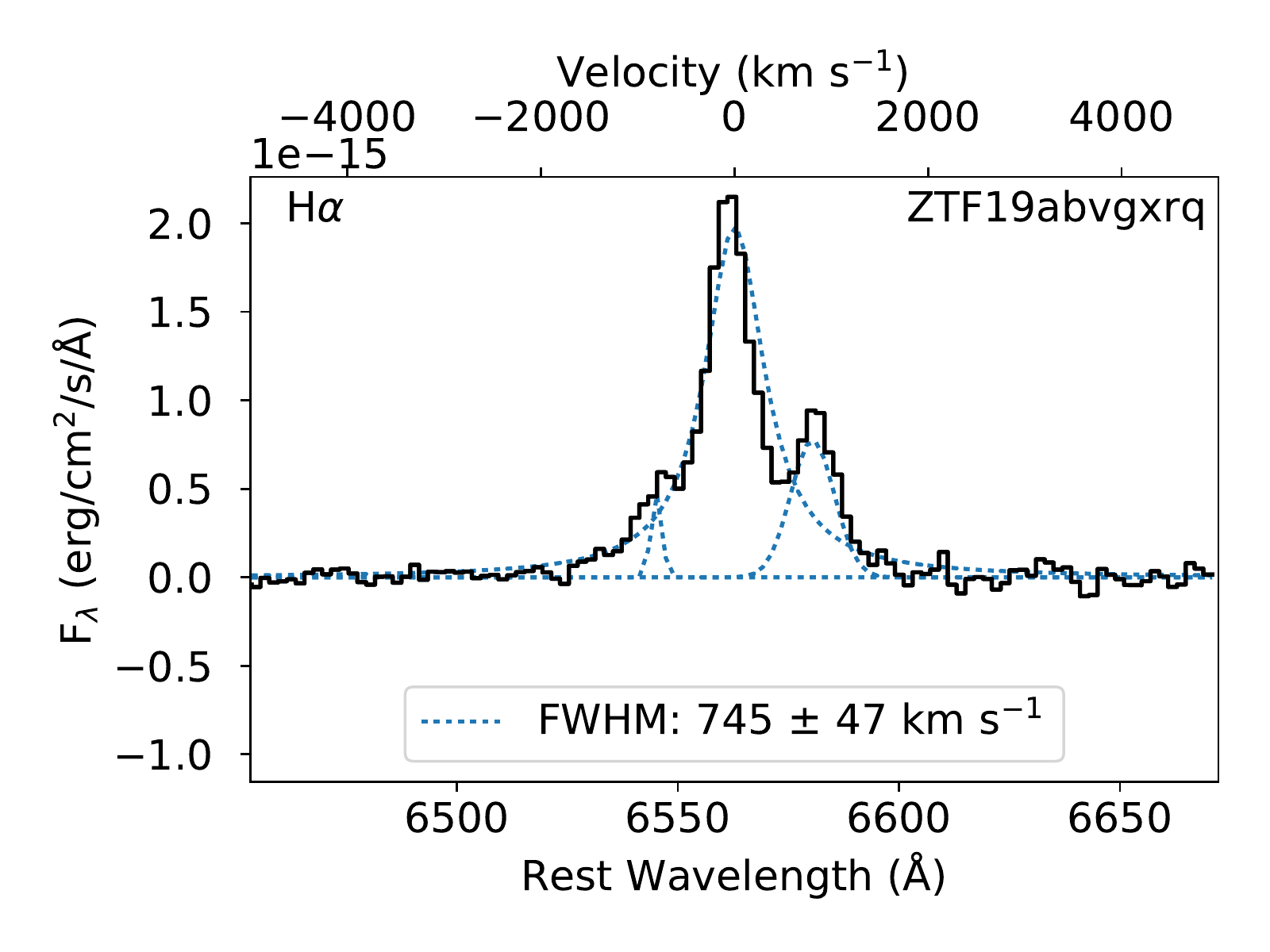}}
  \subfloat[H$\beta$ \stannis~0d]{\label{fig:linefit_stannis_hb}\includegraphics[scale=0.33]{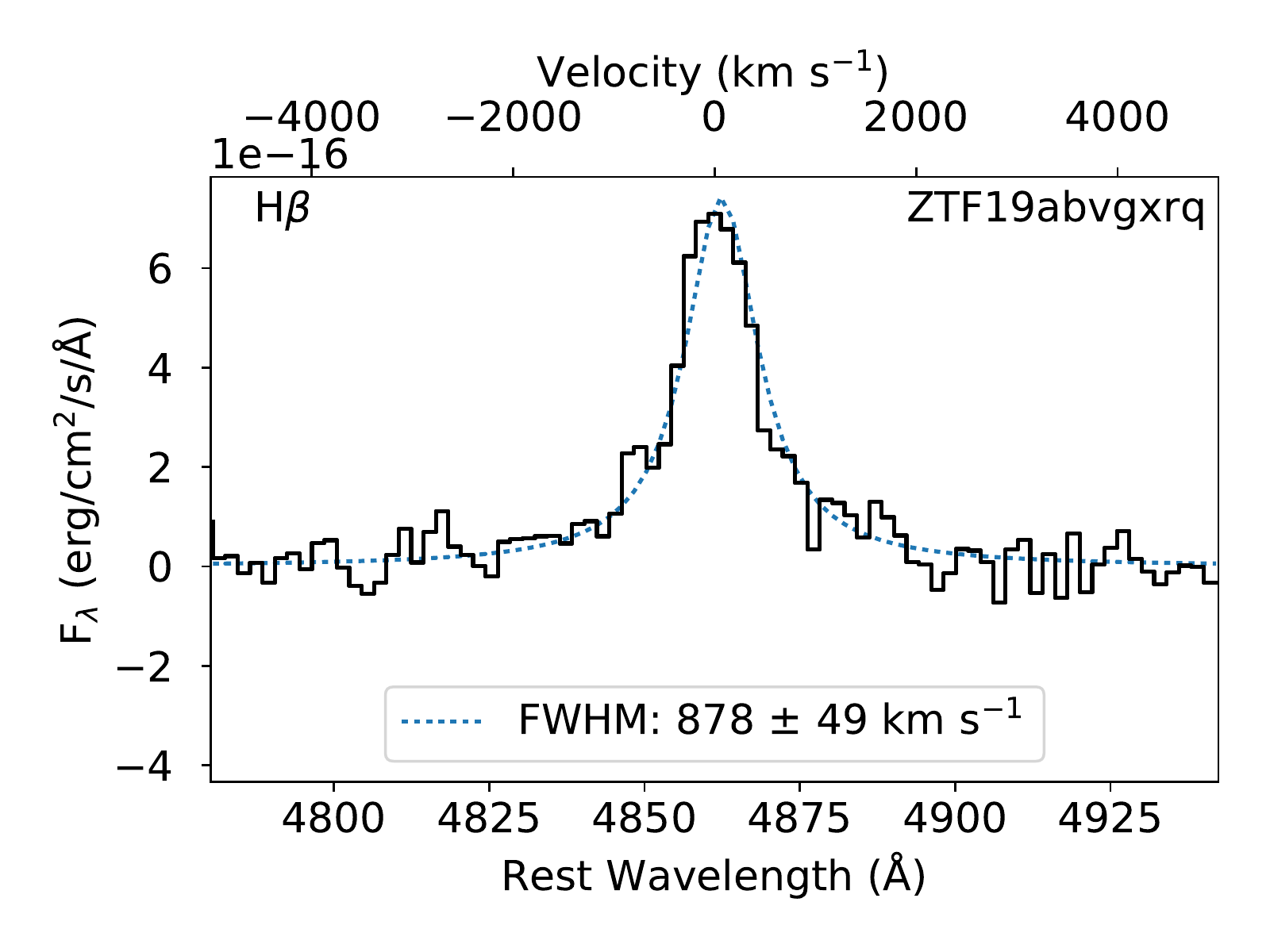}}
  \\
  \subfloat[H$\alpha$ \tywin~$+8$d]{\label{fig:linefit_tywin_ha}\includegraphics[scale=0.33]{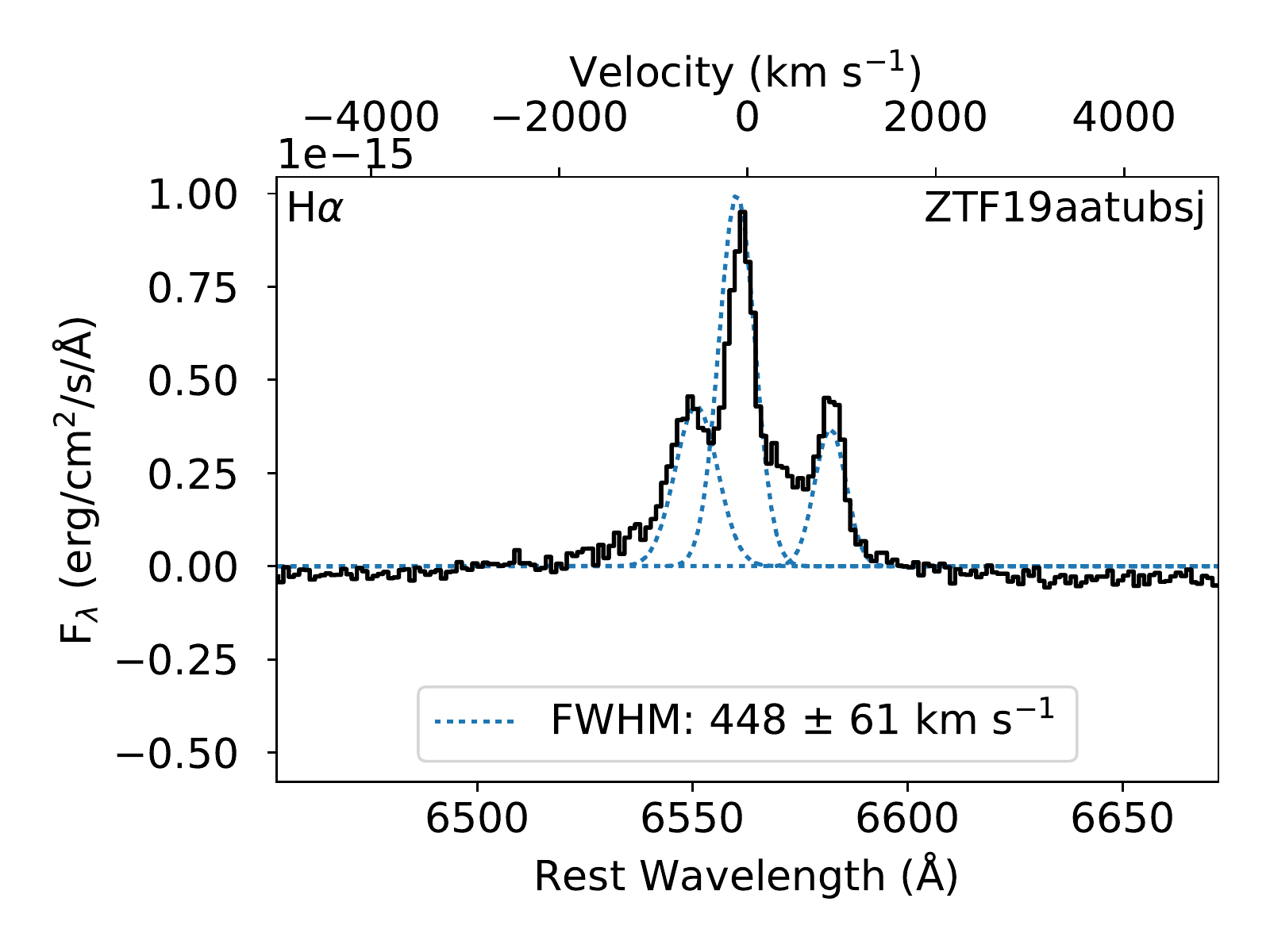}}
  \subfloat[H$\beta$ \tywin~$+$8d]{\label{fig:linefit_tywin_hb}\includegraphics[scale=0.33]{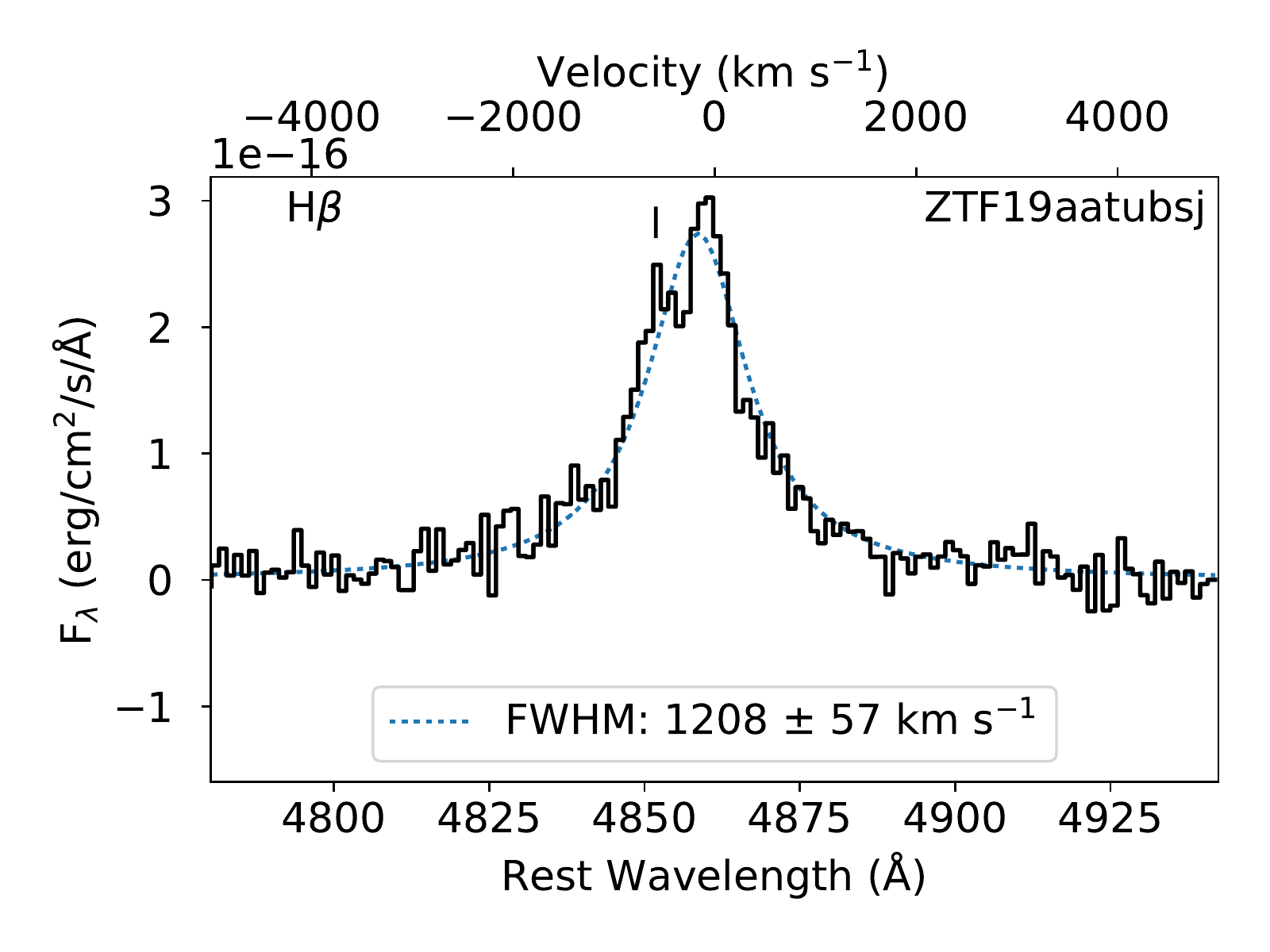}}
  \\
  \subfloat[H$\alpha$ \selyse~$-$13y]{\label{fig:linefit_selyse_ha}\includegraphics[scale=0.33]{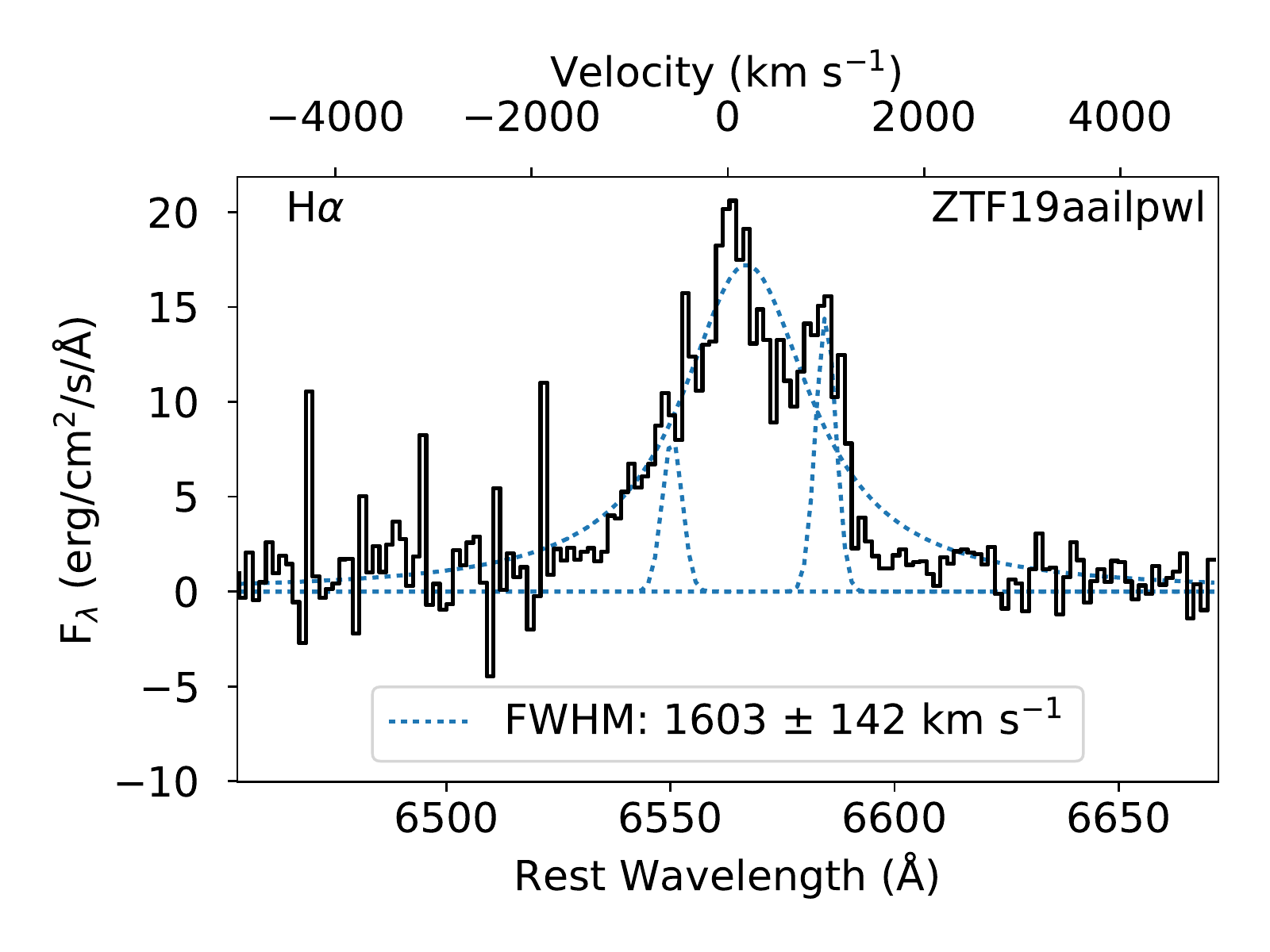}}
  \subfloat[H$\beta$ \selyse~$-$13y]{\label{fig:linefit_selyse_hb}\includegraphics[scale=0.33]{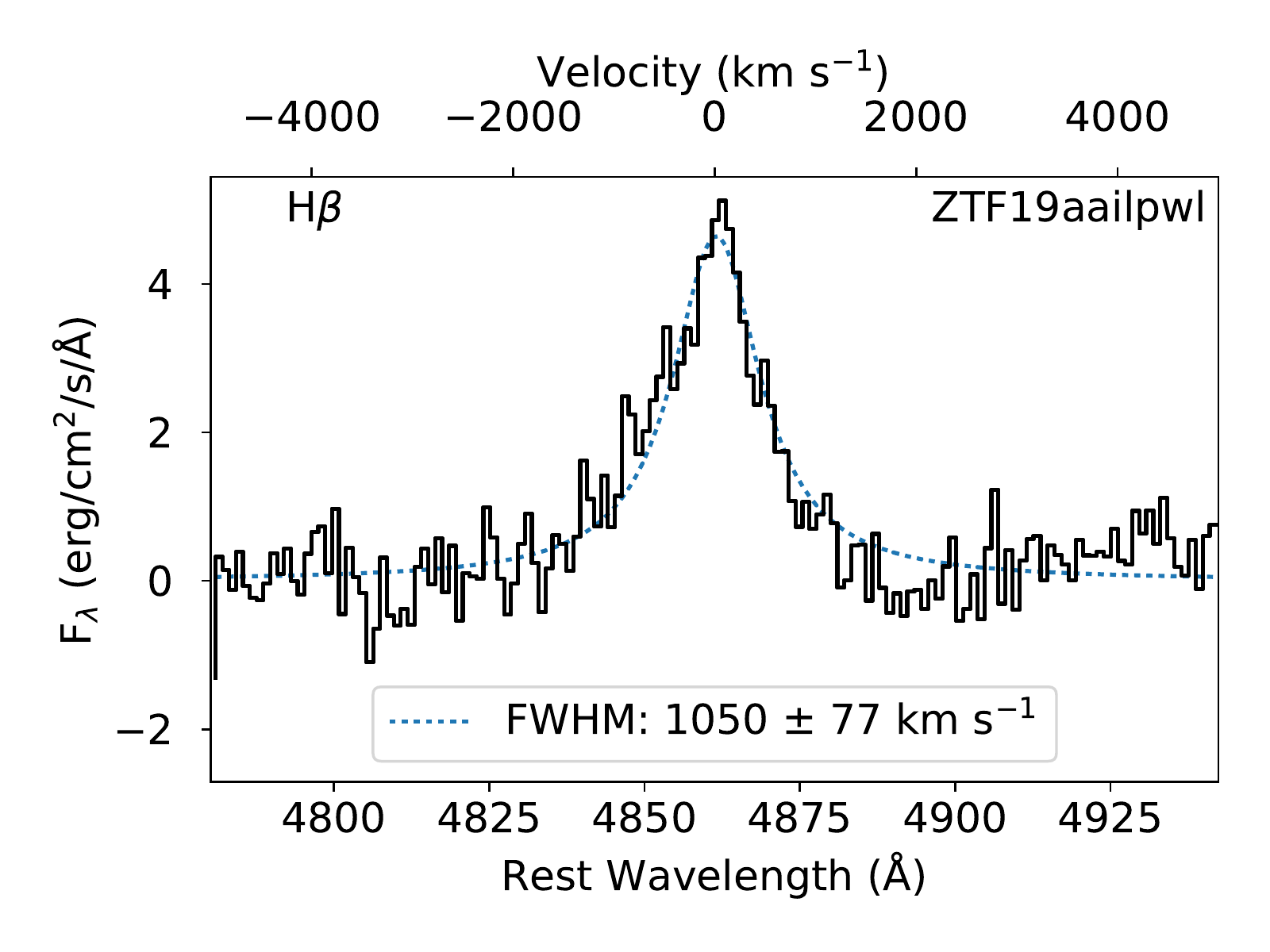}}
   \\
  \subfloat[H$\alpha$ \ciri~$+$84d]{\label{fig:linefit_ciri_ha}\includegraphics[scale=0.33]{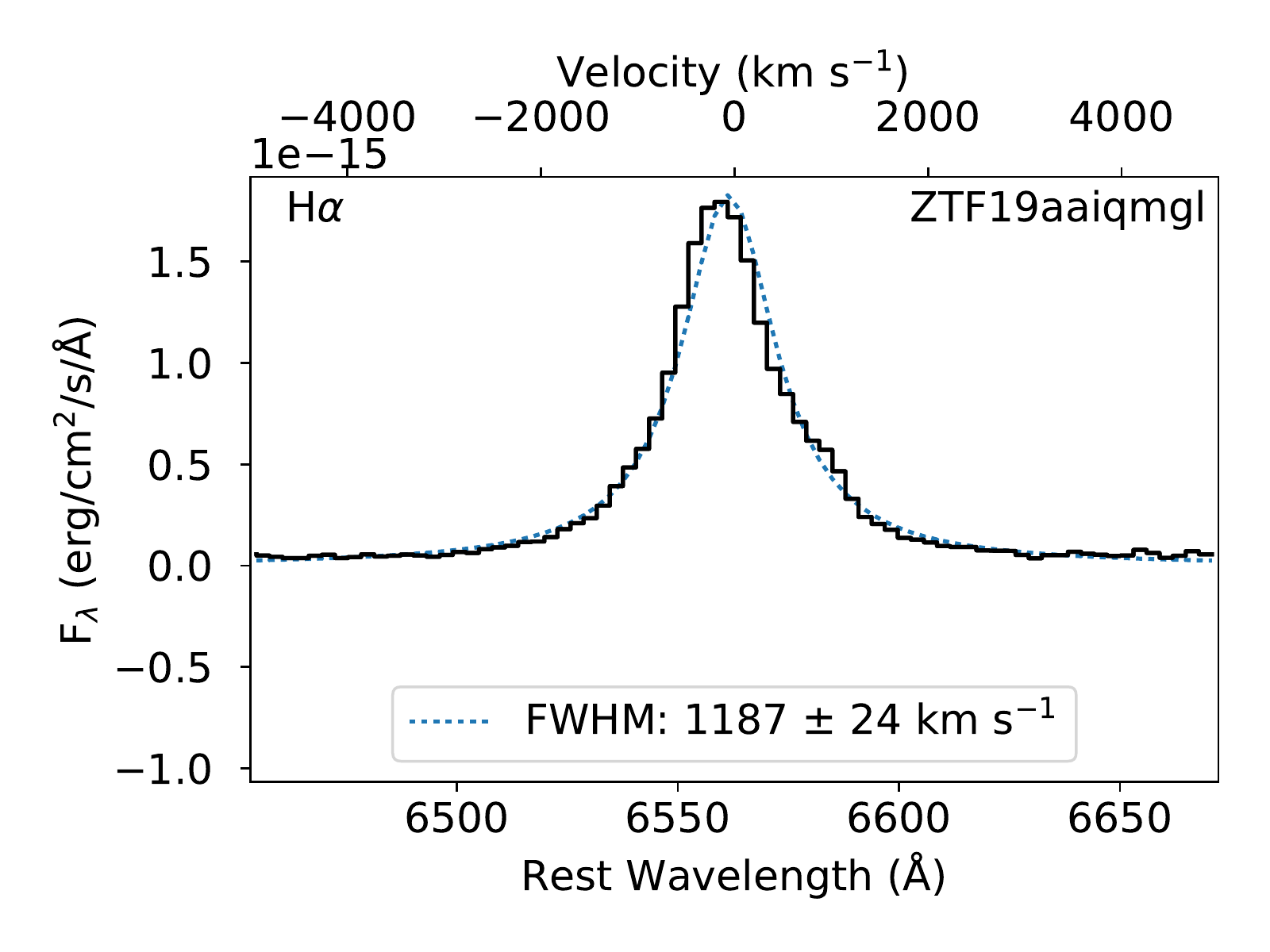}}
  \subfloat[H$\beta$ \ciri~$+$84d]{\label{fig:linefit_ciri_hb}\includegraphics[scale=0.33]{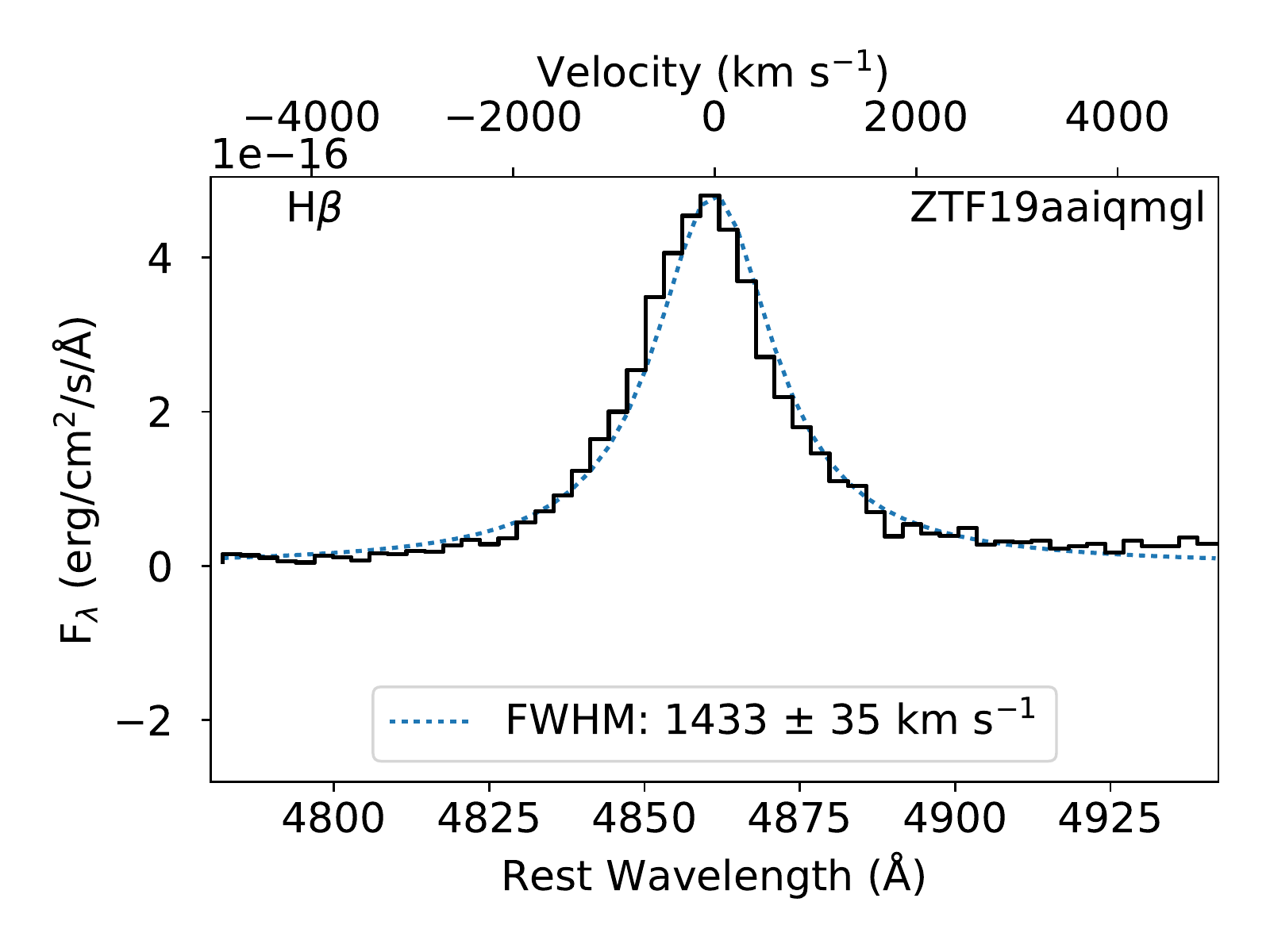}}
   \\
  \subfloat[H$\alpha$ \renly~$+$8d]{\label{fig:linefit_renly_ha}\includegraphics[scale=0.33]{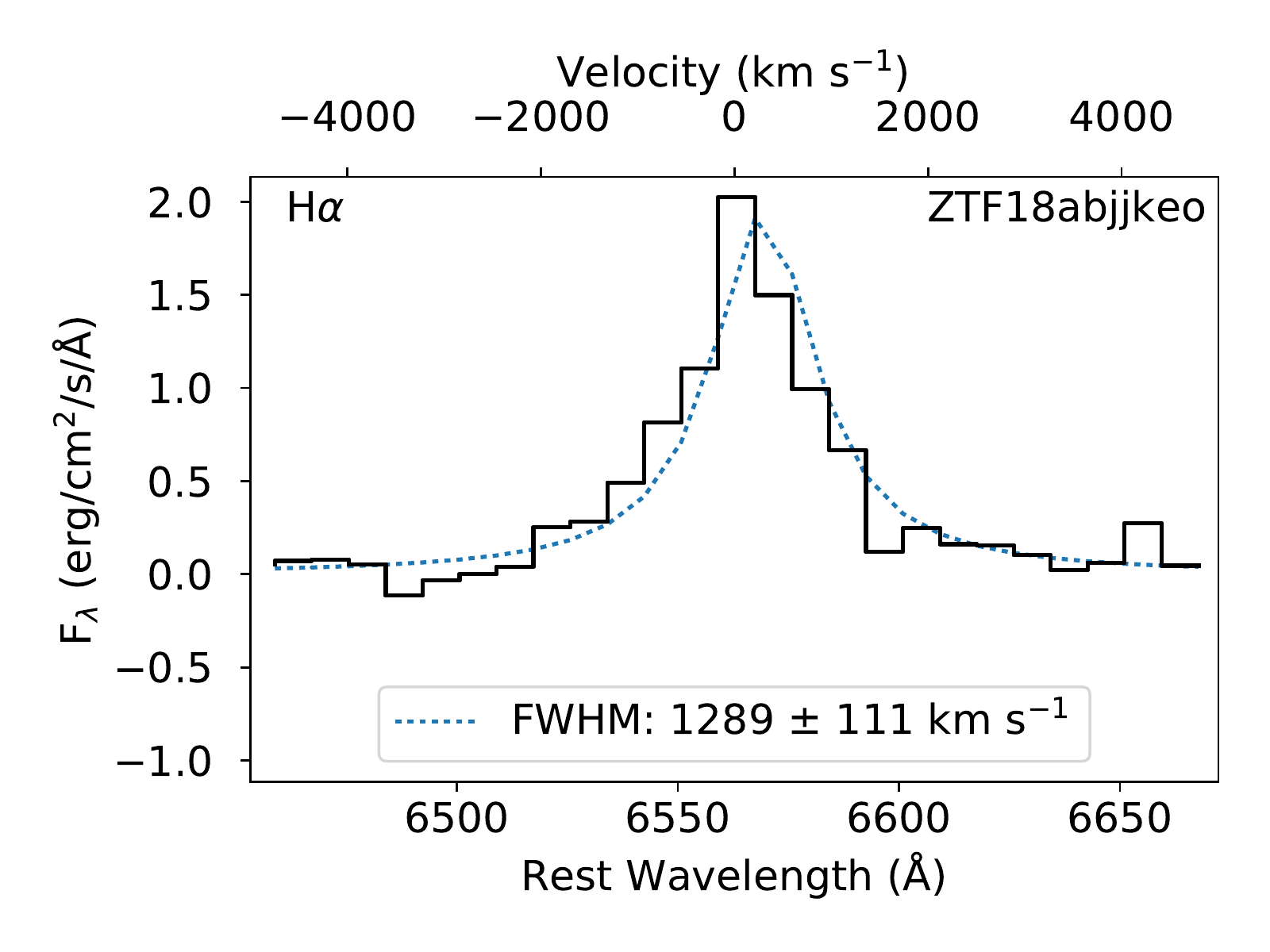}}
  \subfloat[H$\beta$ \renly~$+$8d]{\label{fig:linefit_renly_hb}\includegraphics[scale=0.33]{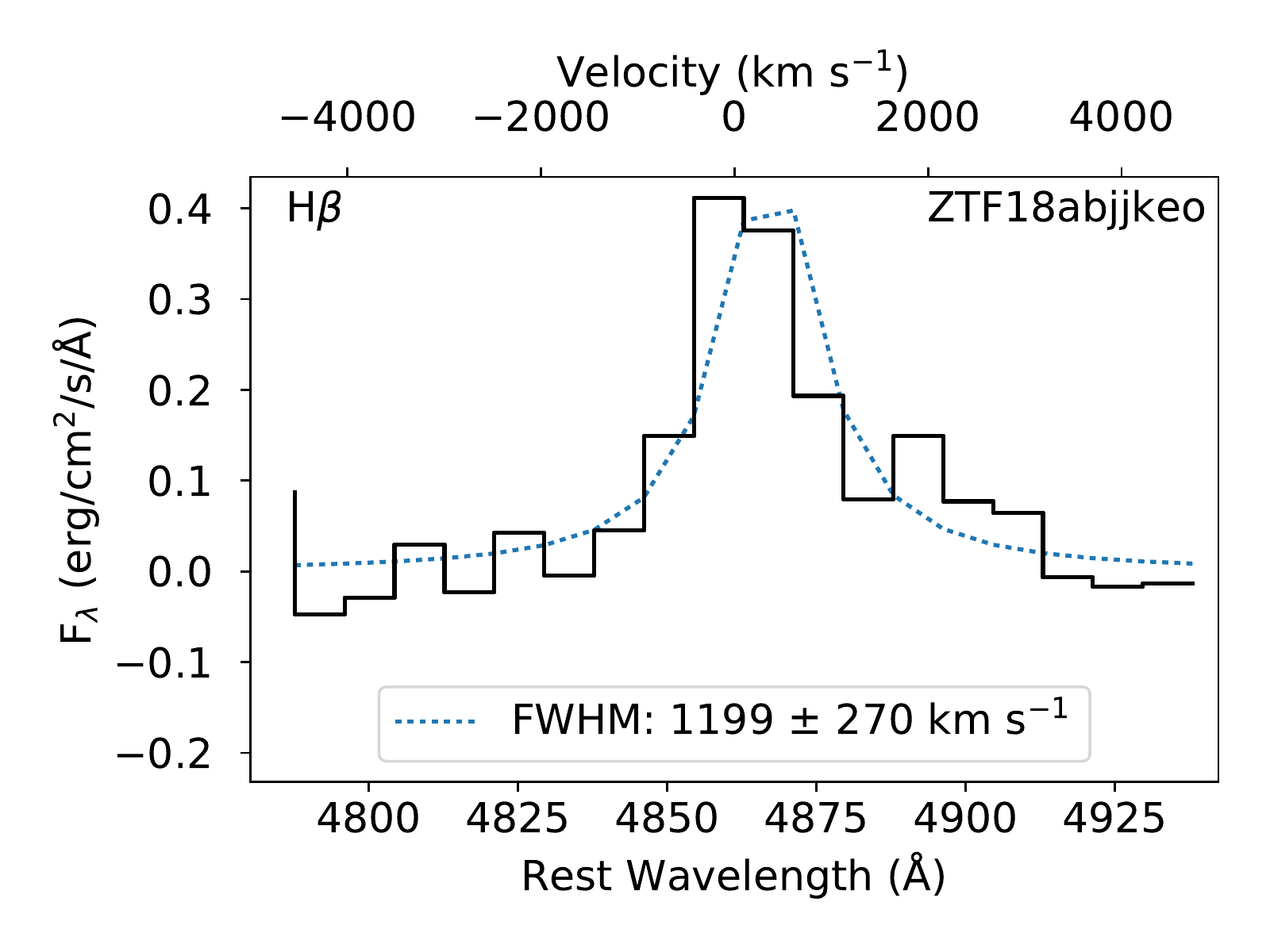}}

\caption{Gaussian fits to the H$\alpha$+[N~II] and H$\beta$ line profiles of all transients in the sample show that their Balmer lines have a FWHM consistent with (and Lorentzian Balmer profiles characteristic of) that of narrow-line Seyfert 1s. The offset blue peak in the H$\beta$ profile of \tywin~is marked by a vertical line.}\label{fig:linefits}

\end{figure*}

From the FWHM of the broad Balmer emission lines, we classified all sources in the sample as NLSy1s.
We fit the H$\alpha$ and H$\beta$ line profiles of the host (when available) and transient spectra of the sample with the non-linear least-squares minimization and curve-fitting routine in the  \texttt{lmfit} Python package. The results of these fits are shown in Figure~\ref{fig:linefits}.  Using a Lorentzian profile for the broad H$\alpha$ component fit provided an improvement of the fit over that of a Gaussian profile, as would be expected based on studies of NLSy1s (e.g. \citealt{Nikolajuk2009}). 

\begin{figure*}[ht!]
\includegraphics[scale=.85]{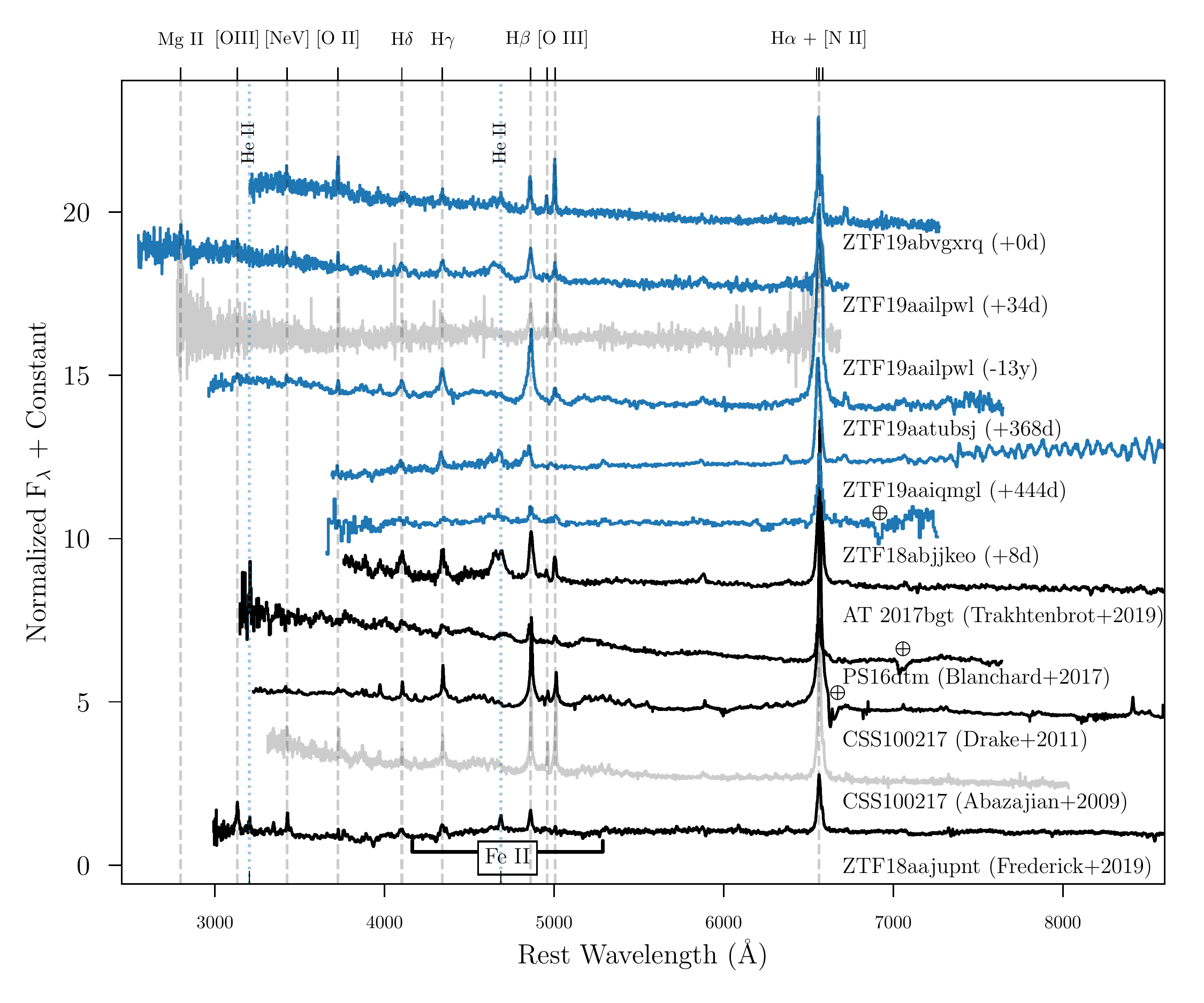}
\caption{Comparison of the ZTF sample of flares (in blue), as well as discovery spectra for the NLSy1-related events from the literature (in black): changing-look LINER \tyrion~\citep{Frederick2019}, TDE in a NLSy1 PS16dtm \citep{Blanchard2017}, SN in a NLSy1 CSS100217 \citep{Drake2011}, and Bowen fluorescent flare AT2017bgt \citep{Trakhtenbrot2019a}, and their pre-event spectra when available (in grey). For \tywin~and \ciri~here and in Figure~\ref{fig:class}, we plot the spectra after continuum fading rather than the discovery spectra, to display the features used in the spectroscopic classification scheme discussed in Section~\ref{sec:class}. \tywin~and \ciri~show offset blue peaks in H$\beta$, and the peak of He~II is offset from 4686 \AA~in \renly.  }
\label{fig:spc}
\end{figure*}

\begin{figure*}[ht!]
\includegraphics[scale=.85]{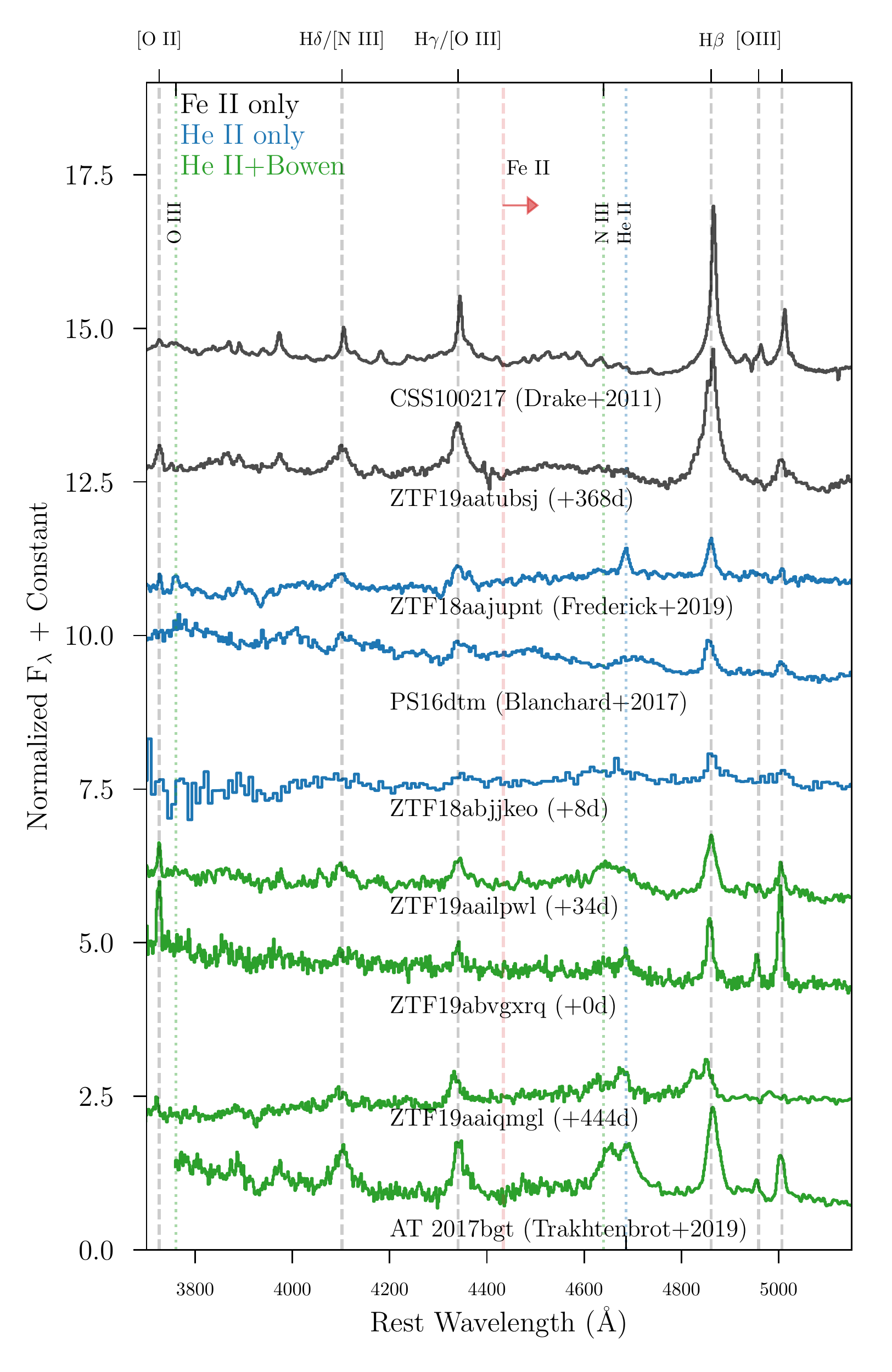}
\caption{Zoom-in on the 4000-5000 \AA~region of Figure~\ref{fig:spc} showing the comparison of the strength of H$\beta$, Fe~II, and [O~III] of the sample with NLSy1-related events in the literature. We color-code the sample and establish categories based on the presence and absence of key emission line features as described in Section~\ref{sec:class}. Blue spectra indicate the presence of He~II, and black spectra indicate transients which displayed Fe~II only, (though we note that PS16dtm showed both features). Those in green display Bowen fluorescence features in addition to He~II, and are most spectroscopically similar to AT2017bgt and the other AGN flares comprising the sample in \citet{Trakhtenbrot2019a}. }
\label{fig:class}
\end{figure*}

We compare the host (when available) and transient spectra of this sample to other transients in NLSy1s in Figures~\ref{fig:spc} (showing the full  wavelength range of the observations) and \ref{fig:class} (rest wavelength $3700-5150$ \AA, showing clearly the He~II, Fe~II, and H$\beta$ line profiles). In Figure~\ref{fig:class} we color-code  the sample (as well as these known NLSy1-related transients in the literature) based on the observational classification scheme we establish in Section~\ref{sec:class}, named after the features discussed in the following sections: ``He~II only'', ``He~II+N~III'', and ``Fe~II only''\footnote{We note that although ``only'' is used in the categorization naming based on the presence of spectral features, all have strong Balmer features.}. 

When compared to the newly discovered flaring events to those in the literature, it is clear that AT2017bgt \citep{Trakhtenbrot2019a} has a much stronger He~II+N~III Bowen fluorescence profile, CSS100217 \citep{Drake2009} has stronger narrow emission lines overall, and  \tyrion~\citep{Frederick2019} has a weaker blue continuum. The presence and strength of Fe~II is uncorrelated with other spectroscopic properties of the transients shown. Of the ZTF sample, the transient spectrum of \tywin~shows the strongest Fe~II complex. However, \tywin~shows no strong He II + Bowen fluorescence features while the others in the ZTF sample do. \tywin~and \ciri~both show offset blue components of H$\beta$. 


\subsubsection{Strong He II profiles in AGN?}
In the discovery paper for transient ASASSN-18jd, \citet{Neustadt2020} emphasized the relatively rare nature of strong He~II emission in AGN in general, noting the exceptions in the \citet{Trakhtenbrot2019a}~observational class of flares as well as the rapid changing-look AGN event \tyrion~\citep{Frederick2019}. A strong He~II line profile is common (but not ubiquitous) in the spectra of TDEs, and they are typically accompanied by Bowen fluorescence features (e.g. \citealt{Blagorodnova2019,vanVelzen2020}). \ciri, \stannis, \selyse~look the most similar to AT2017bgt spectroscopically. They are spectroscopically classified as ``He~II+N~III"-type flares in Figure~\ref{fig:class}. 

\subsubsection{The Fe~II complex}
A strong Fe~II line complex (blueward and redward of H$\beta$+[O~III] in optical spectra, between 4434 \AA~and 5450 \AA) is a distinguishing feature of NLSy1 galaxies. Reverberation mapping studies of AGN show that the line complex emitting region is measured farther than the Balmer line emitting region (e.g. \citealt{Barth2013,Rafter2013}). The Fe~II complex seen in PS16dtm was interpreted as evidence of the system being a NLSy1 prior to the onset of the flare. \css~also displayed a strong Fe~II complex and was interpreted as a SN in a NLSy1 \citep{Drake2011}. TDE AT2018fyk also showed low ionization lines including an Fe~II (37,38) emission multiplet emerging for 45 days during the tidal disruption event, and forms a class of Fe-rich TDEs along with ASASSN–15oi and PTF–09ge \citep{Wevers2019}. Therefore, this feature may indicate the presence of an AGN, but is not always useful in determining the nature of a particular AGN-related flare. For two of the transients in this sample, whether or not the Fe~II complex can be seen in optical spectra depends on the phase and the continuum brightness of the transient --- for \tywin~it was not observed for 368 days, and for \ciri~it became no longer visible during the second rise 444 days after the initial spectrum was taken.





\subsection{X-rays}

\begin{figure*}[ht!]
\centering
\subfloat[\stannis]{\label{fig:xrt_stannis}\includegraphics[scale=0.42]{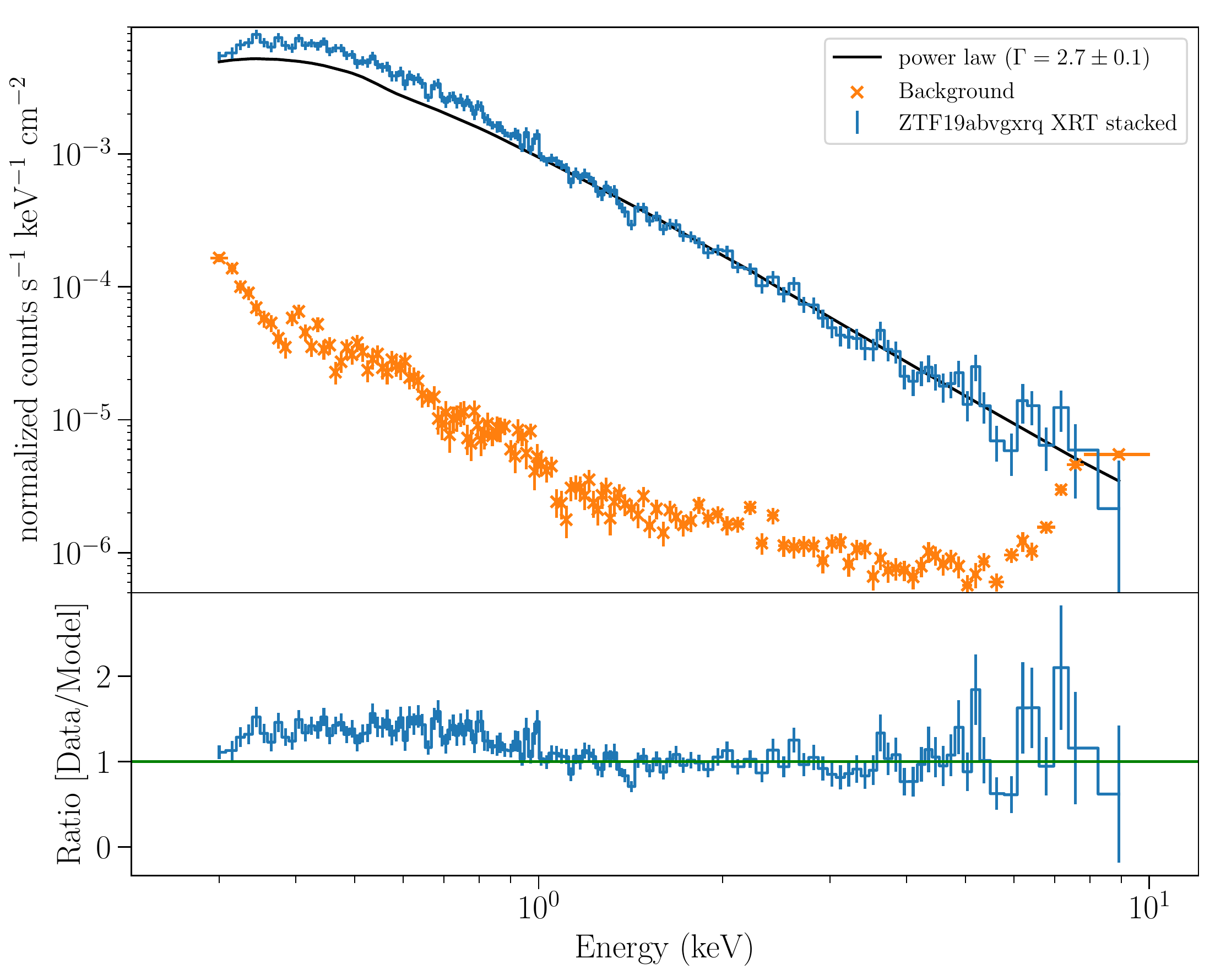}} 
\subfloat[\ciri]{\label{fig:xrt_ciri}\includegraphics[scale=0.42]{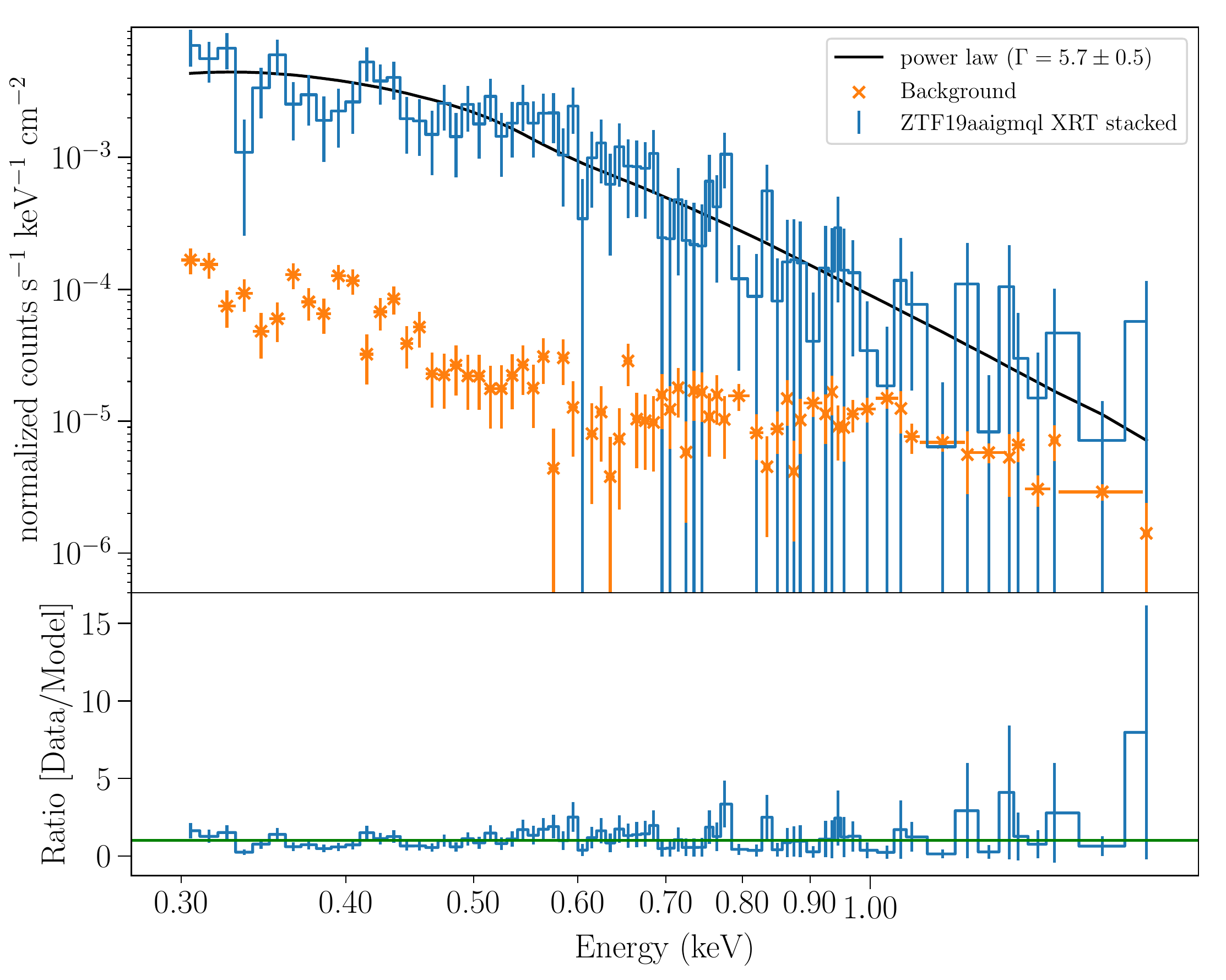}} 
\caption{Left panel: An absorbed power law fit~and ratio residuals  to the $\sim$100 ks stacked {\it Swift XRT} spectrum of \stannis~(spectral index $\Gamma=2.7\pm0.1$). Right panel: The $\sim$4 ks stacked spectrum of \ciri~($\Gamma=5.7\pm0.5$).} 
\label{fig:xrt}
\end{figure*}

There are only two significantly X-ray detected transients in the sample: \stannis~and \ciri. We show their X-ray spectra in Figure~\ref{fig:xrt} fit to power law models. The third, \selyse, was only detected in one epoch and not at a level that allowed for the signal-to-noise necessary for a spectrum.

The X-ray spectrum of \stannis~is measured by {\it Swift XRT} with a power law index of $\Gamma=2.99\pm0.02$, typical of the strong soft X-ray excess observed below 1 keV in NLSy1s \citep[$\overline{\Gamma}=2.8\pm0.9$;][] {Boller1996,Forster1996,Molthagen1998,Rakshit2017}. 
The spectrum of \stannis~could also be explained by a 150 eV blackbody with a $\Gamma=2$ power law component and no intrinsic absorption \citep{Kara2019ATel}. We note that the soft excess observed in NLSy1s can mimick the blackbody temperatures expected for TDEs (e.g. \citealt{Boller1996}). 

The spectral index of \stannis~($\Gamma\sim3$) was similar to that of AT2018fyk, interpreted as a TDE with late-time disk formation \citep{Wevers2019}, as well as \tyrion, interpreted as a changing-look LINER ``turning-on'' into a NLSy1 \citep{Frederick2019}. The X-ray spectral index of \ciri~was quite high even with regard to these events, with $\Gamma\sim4-6$.

\subsection{Black Hole Masses} \label{sec:mbh}

We measured the black hole masses of the sample using two different methods, each with important caveats: The virial mass method, which may systematically underestimate BH masses for NLSy1s, and the host galaxy luminosity, which may be contaminated by the presence of an AGN. The $M_{\rm{BH}}$ calculated from the host galaxy luminosity is $M_{\rm{BH},M_r}=-0.5M_{r,\rm{host}} - 2.96$ following \citet{McLure2002}, and the standard virial method (e.g. \citealt{Shen2011}) was employed to obtain the virial black hole masses from FWHM H$\beta$ reported in Table~\ref{tab:mbh}. The transient Eddington ratio estimates depend on the BH masses ($M_{\rm{BH}}$) as $L_{\rm{Edd}} = 1.3\times 10^{38}~(M_{\rm{BH}}/M_\odot)$ erg s$^{-1}$, For each transient in the sample, we report a range of Eddington ratios in Table~\ref{tab:mbh} bracketed by the Eddington ratio measured assuming the virial mass estimate for the BH mass, and the Eddington ratio measured assuming BH mass derived from the host galaxy luminosity. The range in BH masses, and therefore Eddington ratios, shown in Table~\ref{tab:mbh} is quite large. We estimate statistical and systematic uncertainties of $0.3 - 0.5$ dex on these mass and Eddington ratio measurements, due to the typical scatter associated with single-epoch mass scaling relationships as well as the unknown BLR geometry (e.g. \citealt{Merloni2015,Runnoe2016,Liu2018,Liu2020}).

\citet{Miller2019} obtained an independent measurement of the BH mass of \stannis. They measured  $M_{\rm{BH}}=3.7\times10^6 M_\odot$ from the observed Chandra X-ray luminosity (this observation is described in more detail in Section~\ref{sec:obs:xray}). This is closer to, but not consistent with, the virial mass estimate, meaning that the transient may not have been accreting near the Eddington limit at the time of the X-ray observation.

\setlength{\tabcolsep}{3pt}

\begin{table*}[ht!]
\caption{Black hole mass measurements of the sample from optical spectra and host galaxy properties.  NLSy1s are typically thought to be lower mass, highly accreting systems, but we show here that the uncertainty in the mass estimates generates significant uncertainty in the estimates of the Eddington ratios (described in Section~\ref{sec:mbh}). $M_{r,\rm{host}}$ is the $r$-band de Vaucouleurs and exponential disk profile model fit magnitude from the SDSS DR14 photometric catalog. The host of \renly~is not in the SDSS footprint, and so we instead use the Pan-STARRS1  $r$-band Kron magnitude of this source  \citep{Chambers2016}.} 
\hspace{-4em}\begin{tabular}{@{}ccccccccccc@{}}
\toprule
Name            & $M_{r,\rm{host}}$ & $\lambda L_{5100A}$      & FWHM$_{H_{\beta}}$ & log $M_{\rm{BH},M_r}$ & log $M_{\rm{BH,vir}}$ & $L/L_{\rm{Edd}}$ \\ \midrule
                & (mag)             & ($10^{43}$ erg s$^{-1}$) & (km s$^{-1}$)      & {[}$M_{\odot}${]}     & {[}$M_{\odot}${]}     &                  \\ \midrule
$\rm{\stannis}$ & -21.36            & 5.00 $\pm$ 0.04          & 878 $\pm$ 49       & 7.7                   & 6.4                   & 0.066-1.5        \\
$\rm{\renly}$   & -20.94                & 2.24 $\pm$ 0.02          & 1199 $\pm$270      & 7.5                     & 6.4                   & 0.048-0.62             \\
$\rm{\selyse}$  & -22.38            & 42.6 $\pm$ 0.8           & 1050 $\pm$ 77$^{\rm a}$     & 8.2                   & 7.2                   & 0.17-1.97        \\
$\rm{\ciri}$    & -20.35            & 0.553 $\pm$ 0.008        & 1433 $\pm$ 35      & 7.2                   & 6.1                   & 0.023-0.29       \\
$\rm{\tywin}$   & -21.51            & 21.9 $\pm$ 0.2           & 1208 $\pm$ 57      & 7.8                   & 7.1                   & 0.24-1.2         \\ \bottomrule
\end{tabular}

 \hspace{4em}  a. The FWHM(H$\beta$) for \selyse~agrees with the measurement in \citet{Rakshit2017} within the error estimates.
\label{tab:mbh}
\end{table*}

\section{Discussion}
\label{sec:disc}

In this section, we rule out possible physical scenarios for each outburst, beginning with core collapse supernovae IIn. We review why the supernova interpretation was quickly ruled out in favor of a supermassive black hole accretion scenario, and discuss how many of the characteristics of the objects are consistent with both NLSy1s and TDEs. 
 We compare the available evidence with other scenarios including TDEs, extreme AGN variability, and binary SMBHs in detail. We also discuss NLSy1 galaxies as the preferential hosts for these and other similar events, and outline a scheme for classifying future events based on the presence of spectral features.

\subsection{``IIn or not IIn?'': Preliminary Observational Classification of the Flare Sample} 
\label{sec:iin}
Identification of the sample presented here occurred with a slew of conflicting preliminary classifications at early times, which we describe below.

\begin{figure*}[ht!]
\includegraphics[scale=.75]{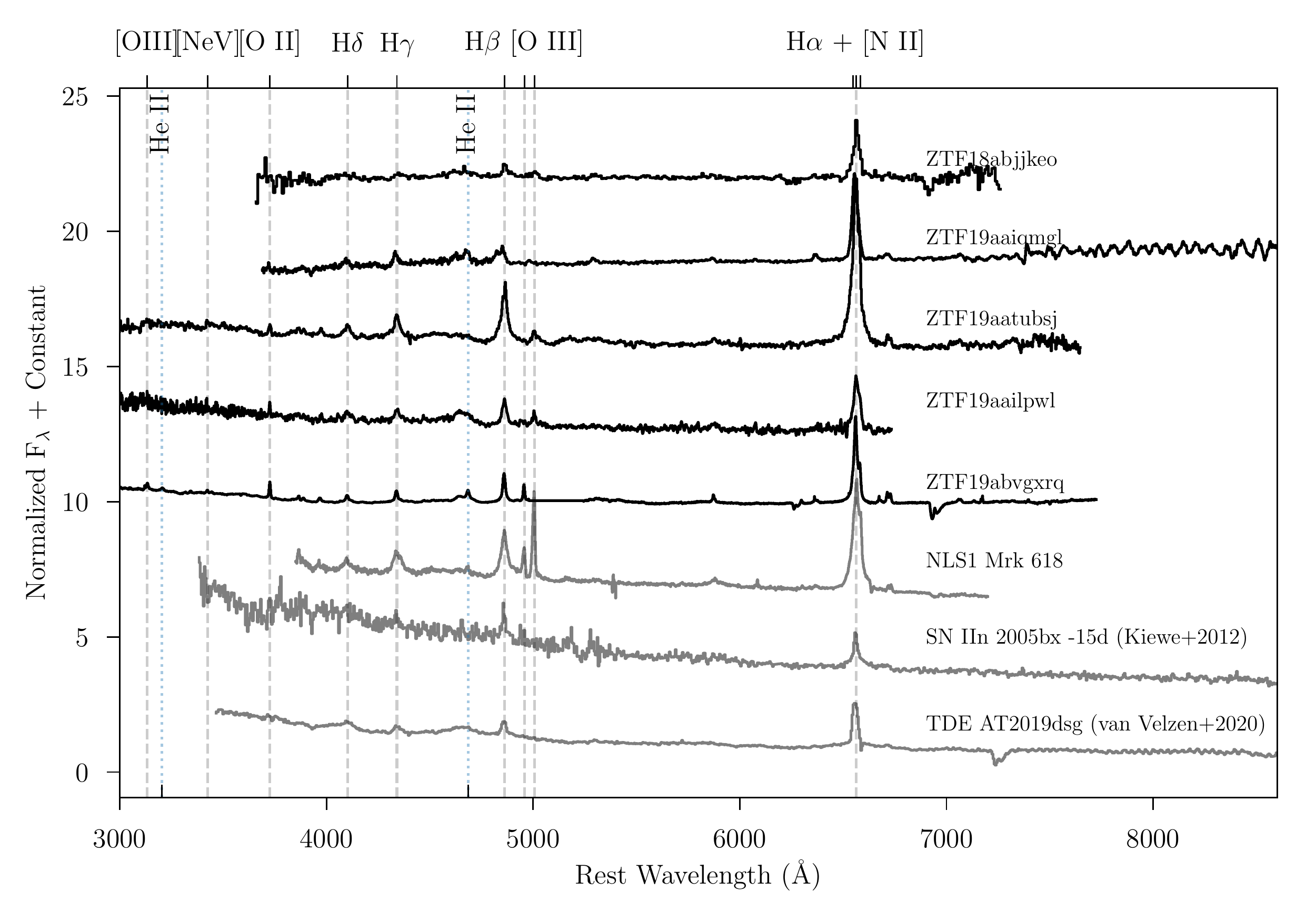} 
\caption{We compare the spectra of the transient sample (in black) to archetypal NLSy1 Mrk 618, as well as a normal Type IIn supernova, SN 2005bx \citep{Kiewe2012}, and AT2019dsg, a normal TDE in a star forming galaxy with Bowen fluorescence features and a coincident neutrino detection \citep{vanVelzen2020,Stein2020}. }
\label{fig:iin_spc}
\end{figure*}

The narrow emission lines in the spectra of some SLSN (Type IIn) are a result of the highly luminous interaction of supernova ejecta from a massive progenitor with dense circumstellar medium. 
Therefore, under special circumstances, nuclear SNe can look spectroscopically very similar to rapid\footnote{With rise times on the order of days to weeks.} flares from NLSy1s in the optical (e.g. \citealt{Moriya2018}). 
The shapes of the light curves of the transients in this sample looked rather like those of such supernovae, in the absence of additional observations. The smoothness of the flares in particular was unique with respect to typical stochastic AGN variability, and made these transients noteworthy for allocation of follow-up resources. 
Therefore, the narrow Balmer features in the spectra of these transients, coupled with their light curve shapes, left uncertainty in their early classifications. They could have been either Type IIn supernovae or NLSy1 AGN, while those with persistent strong He~II~$\lambda$4686 features in their spectra looked similar to that of TDEs. 
To illustrate this, Figure~\ref{fig:iin_spc} shows spectra of the sample alongside a Type IIn SN as well as a TDE with Bowen fluorescence features. 
Additional follow-up observations in the UV/X-rays helped distinguish this sample of transients from SNe. 

\subsection{A Preponderance of Rapid Optical Transients in Narrow-line Seyfert 1 Host Galaxies}
\label{sec:nls1s}
In the Analysis section (\textsection\ref{sec:analysis}), we compared our sample to data from nuclear transients in the literature that happened to be hosted in NLSy1 galaxies. In this and the next sections, we discuss NLSy1s as an interesting AGN subtype, and  observationally classify and link these events to one another on the basis of their shared host properties.

The narrower broad-line Balmer profiles and high amplitude variability, (especially in the X-rays, e.g. \citealt{Pogge2000,Frederick2018}) in NLSy1s may be evidence of smaller black hole masses in these systems (5 $<$ log($M_{\rm{BH}}[M_\odot]$) $<$ 8; e.g. \citealt{Mathur2001}), and/or higher observed accretion rates \citep{Pounds1995,Wang1996,Marconi2008,Grupe2010,Xu2012}. The virial masses derived from spectral measurements of the population may also be explained with geometrical effects, when interpreted as the classic broad-line AGN seen along a lower inclination angle between the broad-line emitting region and the line of sight \citep{Decarli2008,Baldi2016,Rakshit2017}.

Studies of NLSy1s typically find them to be highly photometrically variable only in the X-rays. 
At optical wavelengths, however, \citet{Klimek2004} found that rapid, high amplitude variability was rare in a sample of 172 observations of NLSy1s across 33 nights. \citet{Ai2010} also found that NLSy1s had systematically lower optical variability amplitudes ($\lesssim 0.2$ mag) than broad-line Seyfert 1s in a sample of 275 AGN at $0.3 < z < 0.8$ in 3 years of SDSS data. 

However, optical flares are not unheard of in NLSy1s (e.g. NGC 4051, \citealt{Guainazzi1998,Uttley1999}). 
\citet{Klimek2004} noted the exception of IRAS 13224-3809, which showed both dramatic X-ray and optical variability on short timescales \citep{Miller2000}. 
Here we describe a number of distinct events, including the \citet{Trakhtenbrot2019a} observational class of optical flares, the ``on'' state of \tyrion, and the host of \dtm~and CSS100217:102913+404220, which were all consistent with NLSy1 related activity.

\citet{Trakhtenbrot2019a} established a new observational class of dramatic AGN flares accompanied by Bowen fluorescence features. 
The events in this class all originated from active black holes that were classified as NLSy1 galaxies by their Balmer FWHMs.
Their optical spectra were unusual for NLSy1s in that they showed strong ``double-peaked'' He~II profiles with contributions from the N~III~$\lambda$4640 Bowen fluorescence feature, indicating the presence of a strong UV ionizing continuum. 
This was consistent with the UV brightness observed in the small sample of objects as well as the steep blue continua in these sources. 
The slow UV and spectral emission line evolution over a period of $\sim$450 days ruled out a TDE, and these were instead interpreted as enhanced accretion onto the SMBH of a pre-existing AGN. 
AT2017bgt was presented as the prototype of these dramatic SMBH UV/optical flares irradiating the BLR. It showed a very slow decrease in optical flux over several months following a relatively shallow ($\sim$0.5 mag) rise to peak over $\sim$80 days from a previous non-variable state. During the transient, the X-rays increased by a factor of $2-3$ from a previous  measurement by ROSAT. The persistence of the UV emission over 500 days distinguished it from SNe, and the extremely intense nature of the UV continuum as well as presence of the Bowen fluorescence features in the optical spectrum distinguished it from CLAGN.  Two other NLSy1s, OGLE17aaj \citep{Gromadzki2019} and ULIRG F01004-2237 \citep{Tadhunter2017} (the latter previously interpreted as a TDE), were retroactively reclassified as belonging to this new observational class of NLSy1s.



We compare with \tyrion, a changing-look AGN which transformed from a LINER galaxy to a NLSy1. It was identified as such primarily based on X-ray and UV spectra. It displayed strong high ionizaiton forbidden (i.e. ``coronal'') emission in the optical and UV spectra, an X-ray flare delayed by 60 days, and showed a late-time $g-r$ color change as it faded slowly over 1 year. This was the only AGN with Balmer lines consistent with a NLSy1 among a new class of ``changing-look LINERs'', including SDSS 1115+0544 \citep{Yan2019}.

PS16dtm (iPTF16ezh/SN 2016ezh) was a near-Eddington but X-ray-quiet nuclear transient with strong Fe~II emission and $T_{\rm{BB}} \sim1.7\times10^{4}$ K. It rose over $\sim$50 days to ``superluminous'' levels (log $L_{\rm{bol}}~[\rm{ergs~s^{-1}}] >  44$) at peak before plateauing twice over $\sim$50 and $\sim$100 days while maintaining a constant blackbody temperature. The event was interpreted as a TDE exciting the BLR in a well studied, spectroscopically-confirmed NLSy1 with $M_{\rm{BH}} \sim 10^6 M_\odot$\citep{Blanchard2017}. X-ray upper limits showed dimming by at least an order of magnitude compared to archival observations, but \citealt{Blanchard2017}  predicted the X-rays would reappear after the obscuring debris (oriented perpendicularly to the accretion disk) had dissipated. 
We show the $V$-band ASASSN photometry for PS16dtm in Figure~\ref{fig:rise} which appears similar in shape and absolute magnitude to \tywin, though longer in duration.

CSS100217:102913+404220 displayed a high state ($M_V = -22.7$ at 45 days post-peak) accompanied by broad H$\alpha$ and was interpreted either as a Type IIn SN \citep{Drake2011} or a TDE \citep{Saxton2018} near the nucleus ($\sim$150 pc) of a NLSy1 in a star forming galaxy. 
It eventually faded back to slightly below its original level after one year, which was interpreted as interacting with and subsequently flushing a portion of the accretion disk.

Similar events are not unheard of in broad-line AGN systems, though they may be comparatively more rare. \citet{Neustadt2020} reported a candidate for such a rapidly flaring event with quasar-like properties, ASASSN 18jd, although continued observations of this transient will be critical for a better understanding of the properties of the host.

\subsection{Observational Classification: The ``Family Tree'' of NLSy1-associated Transients} \label{sec:class}

In Table~\ref{tab:properties}, we use this sample to motivate a framework for quickly classifying similarly ambiguous flaring events. We investigate the following:
\begin{itemize}
\item AGN/NLSy1 characteristics (an empirical $W1-W2$ WISE color cutoff from \citealt{Stern2012,Assef2013}, which is comparable between NLSy1s and broad-line Seyfert 1s; \citealt{Chen2017}; a strong Fe~II complex; narrow Balmer emission; and [O~III]/H$\beta < 3$; \citealt{Rakshit2017}), 
\item TDE characteristics  (host black hole mass below the Hills mass ($\sim10^8 M_{\rm{\odot}}$), and a lack of cooling or significant rebrightening), 
\item X-ray properties (the presence of which can occur in both AGN and TDEs, but are less likely in the SN scenario). 
\end{itemize}
We apply these criteria in Table~\ref{tab:properties} and color code them as blue or green based on whether they favor the TDE or AGN scenario, respectively (as the SN scenario has been ruled out in Section~\ref{sec:iin}). 
The spectroscopic class, based on the presence of N~III Bowen fluorecence features, Fe II, and/or He~II~$\lambda$4686, which can occur in both TDEs as well as flaring NLSy1s, is then interpreted in the context of one of these scenarios. 
Based on this table, we confirm the interpretations for three of the four NLSy1-associated transients reported in the literature, except for \css~for which we favor the AGN scenario over the SN interpretation.

Summarized briefly: We expect transients with strong Fe~II complexes are most likely associated with AGN, those with very steep soft X-ray spectra ($\Gamma > 5$) and no intrinsic absorption are most likely associated with TDEs, and those with strong Bowen fluorescence profiles and slow UV and spectral evolution are likely associated with enhanced accretion onto supermassive black holes from a pre-existing accretion disk. 
The timing of a mid infrared flare may also help to distinguish between an AGN and a TDE --- if it precedes the optical, it is likely associated with AGN variability, but if it follows as an echo, it may be associated with a TDE \citep{vanVelzen2016}.

\citet{vanVelzen2020} established a spectroscopic classification scheme for the sample of TDEs discovered during the first half of the ZTF survey, distinguishing those with and without He~II in a single epoch. About half of the TDEs in that sample were ``H-only'', and only one was ``He-only''. They found that  higher density conditions were likely for the rest of the TDEs which had H and He lines, as well as Bowen features. 

For the flaring NLSy1 sample presented here, we establish the following spectroscopic classes to describe each of the transients based on the presence or absence emission features crucial to their physical interpretations:
\begin{enumerate}
\item ``He~II only'', 
\item ``He~II+N~III'', and 
\item ``Fe~II only'', 
\end{enumerate}
and we propose the following naming convention for these classes: ``NLSy1-HeII'', ``NLSy1-HeII+NIII'', and ``NLSy1-FeII''.\footnote{We note that although hydrogen features are not explicitly named in this feature classification scheme, all spectra of the transients show resolved narrow ($1000 < $ FHWM $< 2000$ km s$^{-1}$) Balmer features (see Section~\ref{sec:anal:spc}).} 

\begin{table*}[ht!]  
\caption{Comparison of the properties of individual objects in the sample (upper table) and NLSy1-related transients in the literature (lower table). "$\checked$" means that property is observed, and "$\times$" indicates that characteristic was not observed. ``UV-bright'' refers to the persistence of UV-brightness, and ``Rebrighten'' refers to a significant recovery of at least half the peak luminosity of the source. Following the convention of Figure~\ref{fig:class}, blue (green) indicates a property associated with the TDE (flaring AGN) scenario. 
} 
\hspace{-13em}\begin{tabular}{|l|c|c|c|c|c|c|c|c|c|c|c|c}
\toprule
Name & \cellcolor[HTML]{bcecf4} log$M_{\rm{BH}}$\textless 8 & \cellcolor[HTML]{bcf4bf}  H$\beta$\textless 2000 & \cellcolor[HTML]{bcf4bf} Fe~II & \cellcolor[HTML]{bcf4bf} [OIII]/H$\beta <3$ &  \cellcolor[HTML]{ bcecf4} $\Delta g-r$ & \cellcolor[HTML]{ bcecf4} UV-bright &  X-ray $\Gamma$ & \cellcolor[HTML]{bcf4bf} W1-W2 & \cellcolor[HTML]{bcf4bf} Re- & Spec. class & Interp.    \\ 

& \cellcolor[HTML]{bcecf4} [$M_{\rm{\odot}}$]  & \cellcolor[HTML]{bcf4bf}  km s$^{-1}$  & \cellcolor[HTML]{bcf4bf}  & \cellcolor[HTML]{bcf4bf} [flux ratio] &  \cellcolor[HTML]{bcecf4} $\sim 0$ mag  & \cellcolor[HTML]{bcecf4}  &   & \cellcolor[HTML]{bcf4bf} $>$0.7 mag$^{\rm{a}}$ & \cellcolor[HTML]{bcf4bf} brighten  &    &   \\ \midrule 

ZTF19abvgxrq &  \cellcolor[HTML]{ bcecf4} $\checked$                    &  \cellcolor[HTML]{bcf4bf} $\checked$                   &  \cellcolor[HTML]{ bcecf4} $\times$    &  \cellcolor[HTML]{bcf4bf} $\checked$ & \cellcolor[HTML]{ bcecf4}  $\checked$   & \cellcolor[HTML]{ bcecf4}   $\checked$ &   \cellcolor[HTML]{bcf4bf} 3 & \cellcolor[HTML]{ bcecf4} $\times$ & \cellcolor[HTML]{bcf4bf}  $\checked$  & HeII+NIII        & \cellcolor[HTML]{bcf4bf} AGN  \\

ZTF19aailpwl  & \cellcolor[HTML]{bcf4bf}   $\times$                   &   \cellcolor[HTML]{bcf4bf} $\checked$                   &  \cellcolor[HTML]{bcf4bf} $\checked$  &  \cellcolor[HTML]{bcf4bf} $\checked$   & \cellcolor[HTML]{bcecf4}  $\checked$  & \cellcolor[HTML]{bcecf4}  $\checked$ &   \cellcolor[HTML]{bcecf4} $\checked^{\rm{b}}$  & \cellcolor[HTML]{bcf4bf} $\checked$  & \cellcolor[HTML]{bcecf4} $\times$  & HeII+NIII              & \cellcolor[HTML]{bcf4bf} AGN     \\

ZTF19aatubsj   & \cellcolor[HTML]{ bcecf4} $\checked$                    & \cellcolor[HTML]{bcf4bf} $\checked$                   & \cellcolor[HTML]{bcf4bf} $\checked$   &  \cellcolor[HTML]{bcf4bf} $\checked$    &   \cellcolor[HTML]{ bcf4bf}   $\times$ &  \cellcolor[HTML]{ bcecf4}   $\checked$    &    \cellcolor[HTML]{ bcecf4}  $\times$   &  \cellcolor[HTML]{ bcecf4}  $\times$  &  \cellcolor[HTML]{ bcecf4}  $\times$  &    FeII         &  \cellcolor[HTML]{bcecf4} TDE  \\

ZTF19aaiqmgl & \cellcolor[HTML]{ bcecf4} $\checked$                      & \cellcolor[HTML]{bcf4bf} $\checked$                     & \cellcolor[HTML]{bcf4bf} $\checked$  & \cellcolor[HTML]{bcf4bf}  $\checked$     & \cellcolor[HTML]{ bcf4bf}   $\times$    &  \cellcolor[HTML]{ bcecf4} $\checked$ &  \cellcolor[HTML]{ bcecf4} 5 &   \cellcolor[HTML]{ bcecf4}  $\times$  & \cellcolor[HTML]{bcf4bf} $\checked$ & HeII+NIII       & \cellcolor[HTML]{bcf4bf}   AGN  \\

ZTF18abjjkeo & \cellcolor[HTML]{ bcecf4} $\checked$                      & \cellcolor[HTML]{bcf4bf} $\checked$                     & \cellcolor[HTML]{ bcecf4} $\times$  &  \cellcolor[HTML]{bcf4bf} $\checked$     & \cellcolor[HTML]{ bcecf4} $\checked$    & - &  - & \cellcolor[HTML]{ bcecf4} $\times$ &  \cellcolor[HTML]{ bcecf4}  $\times$ & HeII       & \cellcolor[HTML]{bcecf4} TDE \\

\midrule
$\rm{\css}$     & \cellcolor[HTML]{ bcecf4} $\checked$                             & \cellcolor[HTML]{bcf4bf} $\checked$                    & \cellcolor[HTML]{bcf4bf} $\checked$     & \cellcolor[HTML]{bcf4bf} $\checked$            &     \cellcolor[HTML]{ bcf4bf}  $\times$        &   \cellcolor[HTML]{ bcecf4} $\checked$ &  \cellcolor[HTML]{bcf4bf} 3     &   \cellcolor[HTML]{bcf4bf}     $\checked$         &  \cellcolor[HTML]{ bcecf4}  $\times$   & FeII           & \cellcolor[HTML]{bcf4bf}  AGN      \\

$\rm{\dtm}$     & \cellcolor[HTML]{ bcecf4} $\checked$                     & \cellcolor[HTML]{bcf4bf} $\checked$                    & \cellcolor[HTML]{bcf4bf} $\checked$     & \cellcolor[HTML]{bcf4bf} $\checked$              &      \cellcolor[HTML]{bcecf4}  $\checked$        &  \cellcolor[HTML]{ bcecf4} $\checked$ &  \cellcolor[HTML]{bcf4bf} $2^{\rm{c}}$ &  \cellcolor[HTML]{ bcecf4}  $\times$ &  \cellcolor[HTML]{bcecf4}   $\times$ & HeII+FeII   & \cellcolor[HTML]{bcecf4} TDE  \\

$\rm{\bgt}$     &  \cellcolor[HTML]{ bcecf4} $\checked$                     &  \cellcolor[HTML]{bcf4bf} $\checked$                    &  \cellcolor[HTML]{bcf4bf} $\checked$     &  \cellcolor[HTML]{bcf4bf} $\checked$             &       \cellcolor[HTML]{ bcecf4}  $\checked$      &  \cellcolor[HTML]{ bcecf4} $\checked$ &   \cellcolor[HTML]{bcf4bf} 2      &      \cellcolor[HTML]{ bcecf4}       $\times$   &  \cellcolor[HTML]{bcf4bf}  $\checked$ & HeII+NIII  & \cellcolor[HTML]{bcf4bf} AGN    \\

ZTF18aajupnt & \cellcolor[HTML]{ bcecf4} $\checked$                     & \cellcolor[HTML]{bcf4bf} $\checked$                    &  \cellcolor[HTML]{ bcecf4}  $\times$       & \cellcolor[HTML]{bcf4bf} $\checked$              &   \cellcolor[HTML]{ bcf4bf}  $\times$   &  \cellcolor[HTML]{ bcecf4} $\checked$ &  \cellcolor[HTML]{bcf4bf} 3        &        \cellcolor[HTML]{ bcecf4}               $\times$       &   \cellcolor[HTML]{ bcecf4} $\times$   & HeII              & \cellcolor[HTML]{bcf4bf}  AGN                                                                             
  \\

\bottomrule

\end{tabular}

a. {We select the less conservative color cut presented in \citet{Stern2012}.} \\
b. {The single low level $XRT$ detection of \selyse~occurred only once throughout the follow-up campaign and was not enough to take a reliable spectral measurement.} \\
c. {The host of \dtm~displayed X-rays only prior to and following the fading of, but not for the duration of, the transient.} 
\label{tab:properties}
\end{table*}


\subsection{Physical Interpretation of the Transient Flares}

In the following section we consolidate all that is known about the relevant properties of each object in the sample, and compare them with the related transients in NLSy1s in the literature, to explore each of the following scenarios: A PS16dtm-like TDE in a NLSy1, A Sharov-21-like microlensing event, a \css-like SN in a NLSy1, and a binary SMBH scenario. 

\subsubsection{Association of the Transients with AGN}

There is evidence that all sources in the sample are associated with AGN rather than distinct explosive events occurring in a normal galaxy. 
Although these outbursts may not necessarily be the result of an intrinsic enhancement in AGN accretion activity, transients with fast-rise/slow-decay (such as those in this sample, along with slow-rise/fast-decay, and symmetric light curve shapes) were well-represented in a sample of 51 AGN flares discovered in CRTS \citep{Graham2017}.

\citet{Rakshit2017} spectroscopically classified the SDSS spectrum of the host galaxy of \selyse~as harboring an AGN NLSy1 $>12$ years prior to the onset of the smoothly flaring transient reported here. 

As evident in Figure~\ref{fig:iin_spc}, the strengths of the Balmer lines in the transient spectra are most consistent with that of a NLSy1. Ne~V~$\lambda$3426, when observable, is typically associated with AGN, and is present in the spectra of these sources. Strong He~II profiles, although somewhat rare in association with normal stochastic AGN variability \citep{Neustadt2020}, have been observed before and interpreted as the signature of a sudden enhancement of accretion (e.g. \citealt{Trakhtenbrot2019a,Frederick2019}).

Persistent X-rays are a likely signature of accretion onto a SMBH rather than a SN. 
A strong soft X-ray excess is characteristic of NLSy1s. However, it is typically accompanied by a hard X-ray continuum component (not present in either X-ray detected transient in this sample), and not nearly as ultra-soft as the X-rays seen in \ciri~($4 \lesssim \Gamma \lesssim 6$), which are slopes more frequently observed in the X-ray spectra of TDEs.  

\subsubsection{The SN Scenario} \label{sec:sn}

It is highly improbable that these flares are the result of normal SN explosions. We observe long-lived $U$-band emission in \tywin, persistent UV emission in all transients in the sample, and strong transient X-ray detections in \stannis~and \ciri. There is also only a small likelihood of a SN in the host galaxy along the line of sight unassociated with the AGN. The strongest evidence against the normal supernova scenario is the persistence of the He~II emission features $\sim10-100$ days after the onset of the flare --- such flash ionization signatures are only visible in supernova spectra at very early times (e.g. \citealt{Khazov2016,Bruch2020}).

At least one of these transients (\tywin) shares a number of properties with \css, which displayed soft X-rays and was interpreted as a SN IIn explosion in an AGN disk. The SN interpretation of \css~was largely based on light curve energetics, which are similar to those of this sample. 
The $g-r$ color change, and the peak magnitude of $-23 \lesssim M_V \lesssim -22$ are very similar in particular between \css~and \tywin. 
Type IIn supernovae can exhibit strong Fe~II lines in late spectra, such as \tywin~did. 

However, in contrast, the light curve evolution differs in that \css~fades at least twice as quickly as \tywin. 
Also, the Fe II complex of \css~was always visible throughout the flare, and \citet{Drake2011} observed  a broad $\sim$3000 km s$^{-1}$  component in H$\alpha$ which got broader with time in subsequent follow-up spectra of \css. 
Strong P Cygni profiles are observed in the optical spectra of SN, and from such profiles we would expect an absence of absorption on the  blue end of the Balmer line profiles, rather than emission as in the spectra of \ciri~and \tywin. 
Therefore, based on this evidence we rule out the SN Type IIn scenario.



\subsubsection{The TDE Scenario}
The Hills mass is the mass for which the tidal Roche radius is equivalent to the gravitational Schwarzschild radius of the black hole, beyond which a star (that would otherwise be tidally pulled apart) is instead left intact as it passes the event horizon \citep{Hills1975}. 
This maximum mass to tidally disrupt a solar-type star just outside the event horizon is $10^8 M_{\rm{\odot}}$. Therefore a SMBH mass significantly above this limit would likely rule out a TDE. 
Of the supermassive black hole masses derived for the host galaxies, only that of \selyse~is inconsistent with a TDE scenario, (although we note that it is consistent within the typical uncertainty for such mass measurements). The range of absolute magnitudes of the flares in this sample ($-23 < M_r < -19$ mag) also tend to be intrinsically brighter at peak than all but one of the ZTF TDEs ($M_r > -20$ mag) reported in \citet{vanVelzen2020b}, AT2018iih ($M_r = -21.5$ mag). 

Similar to TDEs PS16dtm and AT2018fyk, \tywin~showed two distinct plateau stages on month-long timescales after fading, with some slight fading in between. Color evolution is rare but not unheard of for TDEs, and the cooling \tywin~shows post-peak is slow, with the transient still detected in the UV at late times as would be expected for a TDE. 
Optical rebrightening following the intial flare has been interpreted as the result of late time disk formation in a number of TDEs (e.g. \citealt{Wevers2019, vanVelzen2019}). However, rebrightening with high amplitudes returning nearly to pre-flare levels such as that seen in \stannis~and \ciri~has neither been observed\footnote{Except in the case of the periodicity of ASASSN 14ko, which was interpreted as a possible repeating partial TDE \citep{Payne2020}.} nor predicted (e.g. \citealt{Chan2019, Chan2020}) from a TDE. In these cases with rebrightening, a TDE is strongly ruled out.

\ciri~and \tywin, like AT2018fyk, only showed Fe~II at certain times during the flare. \ciri~only displayed Fe~II during its first peak, and in \tywin, the Fe~II complex got more visible as the transient faded. \tywin~is the only transient in the sample with a lack of He~II features in its spectra. Within the \citet{vanVelzen2020} spectral classification scheme for optical TDEs, \tywin~would be a H-only TDE, with the Fe~II complex attributed to the NLSy1 host. It is important to note that the transients with blue horn features in H$\beta$, \ciri~and \tywin, may be signatures of wind ejecta with a velocity distinct from the AGN. 

Enhanced N~III lines such as that seen in the NLSy1-HeII+NIII spectroscopic class (\stannis, \selyse, and \ciri) are a prediction of TDEs in AGN when compared to the host spectrum  \citep{Kochanek2016,Liu2018b,Gallegos-Garcia2018}. Unfortunately, a pre-flare spectrum was only available to test this for \selyse~(Figure~\ref{fig:spc}). 

Many properties of the hosts do not align with what we expect from AGN. The WISE colors, for example, span a broad range of $0.06 - 0.98$ mag (Table~\ref{tab:properties}). 
The IR flare associated with \ciri~could be interpreted as a dust echo, similar to those seen in a number of TDEs \citep{vanVelzen2016}. A host-subtracted SED fit to the {\it Swift} photometry of \ciri~gives a blackbody temperature consistent with that of known TDEs, $10^{4.25}$ K.

The X-ray variability of TDEs can vary erratically during a flare (e.g. \citealt{Wevers2019,vanVelzen2020}). Although soft X-ray excesses with $\Gamma\sim3$ are characteristic of NLSy1s, \ciri~displays an X-ray power law index much higher than typically seen in NLSy1s, and more characteristic of the extremely soft X-ray spectra observed in TDEs. 

Based on the combination of properties shown in Table\ref{tab:properties}, we conclude that two of the flares, \renly~and \tywin, are better explained as TDEs than AGN flares, although the interpretation is not clear-cut.  However, if we assume that their spectra are a combination of the host NLSy1 galaxy and the transient line emission from the TDE, then given their spectral classes given here of NLSy1-FeII and NLSy1-HeII, respectively, then the TDE spectra themselves, in these NLSy1 galaxies, would have to be of the TDE subclasses of TDE-H (H only lines) and TDE-He (He II only lines), respectively, in order to match the observed spectra.

\subsubsection{The Extreme AGN Variability Scenario} 
\citet{Graham2017} presented a sample of quasars displaying extreme variability in CRTS. Some had similar profiles and amplitudes (rising by 2$-$2.5 mag) but longer timescales (500-1000 days) compared to the flares presented here, For example, J002748-055559 rose by nearly $\sim$2 mag compared to the steady level it maintained for several years prior. 

The optical spectra of the transients in the sample presented here belonging to the ``NLSy1-HeII+NIII'' spectroscopic class, as well as the UV brightness of the sample, are consistent with the properties of the observational class of flares with  Bowen fluorescence established in \citet{Trakhtenbrot2019b}. 
However, all of the transients presented here have faster fading timescales than AT2017bgt. \citet{Trakhtenbrot2019b} stated that the fade timescale of AT2017bgt was longer than expected for a TDE. However, we note that at least one TDE in the \citet{vanVelzen2020} sample (that also displayed Bowen fluorescence features) was observed to fade over nearly 15 months,

\subsubsection{The Gravitational Microlensing Scenario}
Flares due to microlensing are expected to be observable in difference imaging surveys with the combined baseline of iPTF and ZTF. 
The rise portions of the light curve shapes of all the transients measured in Section~\ref{sec:timescales} being well-fit by quadratics is consistent with a lensing event, however, all but \renly~have a longer decay with respect to the initial rise. 
Microlensing by multiple foreground sources can give rise to a symmetric (with respect to the fade) double peak  with a dip in the middle of the optical light curve such as that seen in \stannis~\citep{Hawkins1998,Hawkins2004,Schmidt2010}. The cuspy shape of the first peak is also characteristic of microlensing light curves. 
\ciri~also showed a second peak in its light curve, but the first peak was a lot more rapid and luminous than the second. To test this scenario in \renly~would require continuing to observe for an additional flare.

The microlensing scenario, however, would not account for the strong transient Bowen fluorescence features that appear only at late times in \ciri, and only at early times in \stannis~(Figure~\ref{fig:heii_spc}). \citet{Meusinger2010} explained a similar event as a background quasar with a UV flare in J004457+4123, also known as Sharov 21, being microlensed by a foreground star in M31.

Microlensing is characteristically achromatic, and therefore would be ruled out by the clear evidence for $g-r$ color change observed in \tywin.





\subsubsection{The SMBH Binary Scenario} 
Variability on the timescales of years due to a binary SMBHB system would require a subparsec separation (e.g. \citealt{Graham2015b}).  
In such a system, two SMBHs induce tidal torques carving out a cavity in the circumbinary accretion disk, and may be surrounded by their own minidisks at sufficient separations. 
The interaction of accretion streams with the cavity could cause an outburst on the approximate timescales seen in this sample, which is dependent on the properties of the system. This phenomenon is seen in simulations of SMBH binaries (e.g. \citealt{Ryan2017,Gold2019}). 

We see evidence of offset narrow Balmer emission lines in the spectra of \tywin~and \ciri, which may indicate a significant separate physical component, although it is unclear what is contributing to those blueshifted velocities.

\clearpage
\section{Conclusions} \label{sec:concl}

We report \num~nuclear flaring events associated with NLSy1s, all serendipitously\footnote{As the \citet{Trakhtenbrot2019a} observational class was established midway through the ZTF survey, we had not been systematically filtering such events when the population became apparent in the nuclear transients alert stream search. } discovered in ZTF. We measured their photometric characteristics (such as light curve shape, $g-r$ color, and rise to peak luminosity, finding a correlation between rise time and absolute magnitude), and spectroscopic properties. We then established groupings of the objects in the sample based on analyses of the months-long follow-up campaigns of these objects. 
Based on observed groupings of the sample, we propose the following naming scheme of spectroscopic classes of such transients for use in future optical surveys:  ``NLSy1-HeII'', ``NLSy1-HeII+NIII'', and ``NLSy1-FeII''. 
We ruled out the possibility that these are Type IIn supernovae occurring in NLSy1 systems. 
Despite the heterogeneity of the sample's properties, two of the flares presented in this work have multiwavelength characteristics which could be consistent with TDEs in NLSy1s (\tywin~and \renly), with spectral classes of NLSy1-FeII and NLSy1-HeII, respectively. 
This is a high TDE rate relative to quiescent galaxies, which are more abundant than NLSy1s. 
The prevalence of TDE candidates in the NLSy1 AGN class could be a natural result of their hosting smaller black holes compared to typical broad-line AGN, and therefore satisfying the Hills mass criterion for an observable TDE.  However, without pre-event spectra and X-ray imaging to isolate the contribution of the putative TDE to the composite NLSy1+TDE emission, flaring due to extreme AGN variability cannot be definitively ruled out. For two in the sample (\stannis~and \ciri), we can rule out the simple TDE scenario from rebrightening in their light curves, and we determine that they, along with \selyse~(which had a pre-flare NLSy1 spectral classification and a black hole mass estimate too large to host a canonical TDE), are likely outbursts related to enhanced accretion in excess of typical AGN variability, and with spectral features we classify as ``NLSy1-HeII+NIII'', and members of the \citet{Trakhtenbrot2019a} class of AGN flares. 

Given this sample, together with the growing number of interesting rapid optical transients associated with NLSy1s we reviewed in the literature, we posed the question of why such environments are observed to preferentially host these outbursts. 
Given the relative fraction of NLSy1s found with respect to other AGN classes in spectroscopic surveys such as SDSS ($\sim$15\%; e.g. \citealt{Zhou2006,Rakshit2017}), there is likely an underlying factor enhancing this rate. 
We suggest four different possible explanations for this enhancement:  
\begin{enumerate} \setlength{\itemsep}{8pt}  \setlength{\parskip}{8pt} 
\item A selection bias due to shorter timescales for lower mass BH systems (like NLSy1s), which are therefore more likely to be captured within the baseline of wide field optical surveys, 
\item A systematic disregard of smooth flares in broad line AGN during transient searches, or 
\item A true intrinsic rate enhancement due to instabilities causing rapid changes in the observable environments or accretion efficiencies of these systems. 
\end{enumerate} 
Follow-up strategies of optical transients in AGN that are similarly ambiguous at early times may stand to benefit from the framework we offer here. We hope this classification scheme will guide real-time predictions for potential future behavior of large amplitude flares in NLSy1s, which are clearly an interesting population for future study. 
The next step will be to perform a systematic study of the variability of NLSy1s detected in ZTF, to assess the completeness and rate of this sample of transients with smoothly flaring light curves, and compare to a sample of broad-line AGN. 
Expanding on the small number of unusual transients associated with NLSy1s not only sheds light on the parameter space in which they reside, but also provides the framework for a decision tree for understanding such outbursts when they are inevitably captured at higher rates in upcoming wide field surveys. This will be imperative to establish in advance of larger and deeper surveys such as ZTF Phase II and the Vera C. Rubin Observatory (formerly known as LSST; \citealt{Ivezic2019}), to which the timescales of these flares are well-suited. 
Continued multiwavelength monitoring of the entire sample will be important to determine the host properties for those with sparse data prior to the transient, and for understanding the evolution and nature of these flares. 

\acknowledgments
We would like to thank A. Dittman for useful comments. 
S. Gezari is supported in part by NSF CAREER grant 1454816. 
We thank C. Barbarino for reducing the Nordic Optical Telescope observation of \tywin. 
We thank R. Foley for contributing the Lick Shane Kast observation of \stannis. 
Based on observations obtained with the Samuel Oschin Telescope 48-inch and the 60-inch Telescope at the Palomar Observatory as part of the Zwicky Transient Facility project. ZTF is supported by the National Science Foundation under Grant No. AST-1440341 and a collaboration including Caltech, IPAC, the Weizmann Institute for Science, the Oskar Klein Center at Stockholm University, the University of Maryland, the University of Washington, Deutsches Elektronen-Synchrotron and Humboldt University, Los Alamos National Laboratories, the TANGO Consortium of Taiwan, the University of Wisconsin at Milwaukee, and Lawrence Berkeley National Laboratories. Operations are conducted by COO, IPAC, and UW. 
This work was supported by the GROWTH project funded by the National Science Foundation under Grant No 1545949. 
The SED Machine is based upon work supported by the National Science Foundation under Grant No. 1106171.  
The ZTF forced-photometry service was funded under the Heising-Simons Foundation Grant No. 12540303 (PI: Graham). 
These results made use of the Discovery Channel Telescope at Lowell Observatory.
Lowell is a private, non-profit institution dedicated to astrophysical research and public appreciation of
astronomy and operates the DCT in partnership with Boston University, the University of Maryland, the
University of Toledo, Northern Arizona University and Yale University. The upgrade of the DeVeny optical spectrograph has been funded by a generous grant from John and Ginger Giovale. 
Based on observations made with the Nordic Optical Telescope, owned in collaboration by the University of Turku and Aarhus University, and operated jointly by Aarhus University, the University of Turku and the University of Oslo, representing Denmark, Finland and Norway, the University of Iceland and Stockholm University at the Observatorio del Roque de los Muchachos, La Palma, Spain, of the Instituto de Astrofisica de Canarias. 
This research has made use of data obtained through the High Energy Astrophysics Science Archive Research Center Online Service, provided by the NASA/Goddard Space Flight Center.  
We acknowledge the use of public data from the {\it Swift} data archive. 


\vspace{5mm}
\facilities{PO:1.2m, PO:1.5m, Hale, Swift(XRT and UVOT), DCT, NOT, Shane, Liverpool:2m}

\software{Pyraf,Lmfit,HEAsoft,PIMMS}

\clearpage

\bibliography{CLAGN}{}
\bibliographystyle{aasjournal}

\clearpage

\appendix

\begin{figure*}[ht!]
  \centering

\subfloat[\stannis]{\label{fig:ipac_lc_stannis}\includegraphics[scale=0.55]{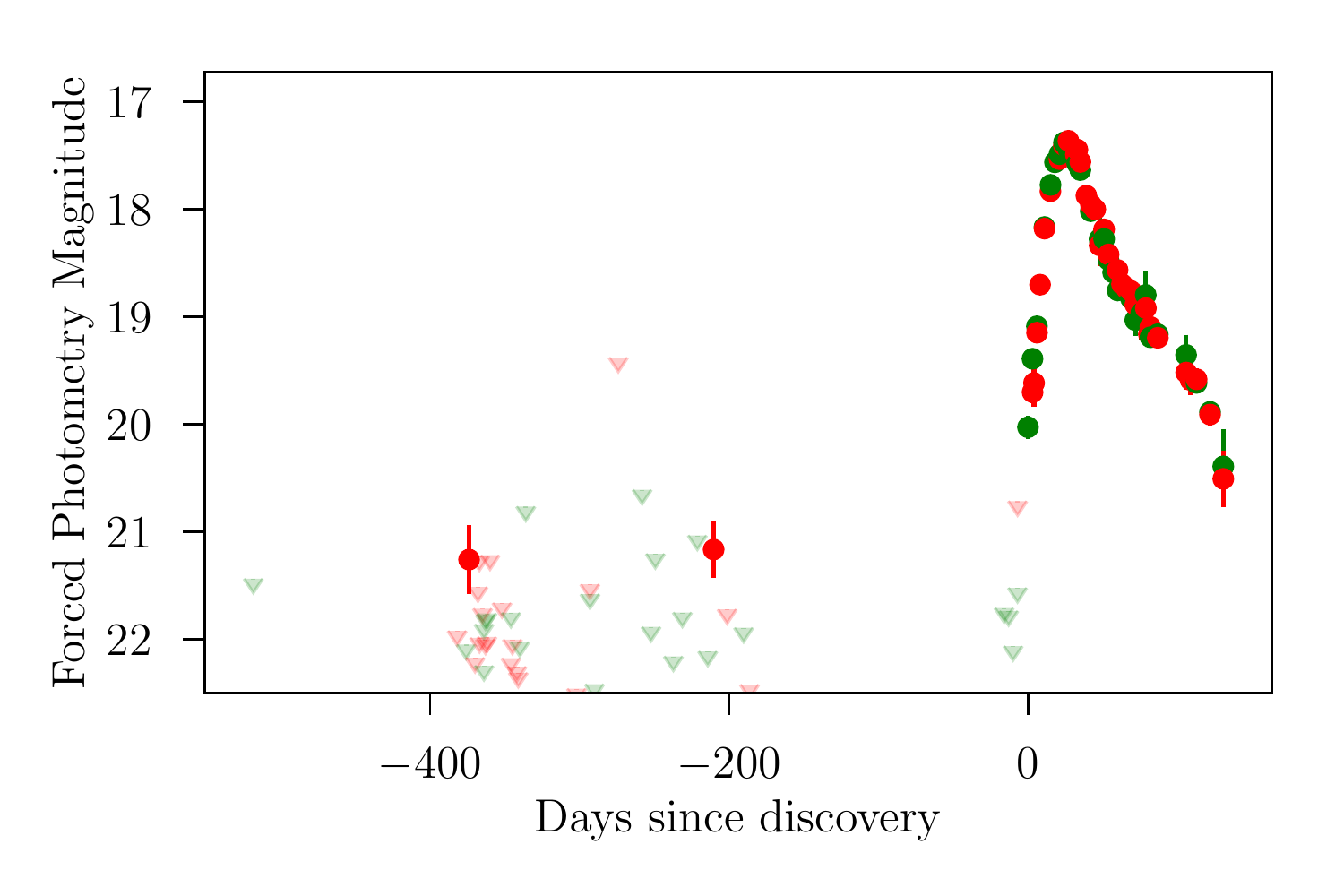}}
  \subfloat[\tywin]{\label{fig:ipac_lc_tywin}\includegraphics[scale=0.55]{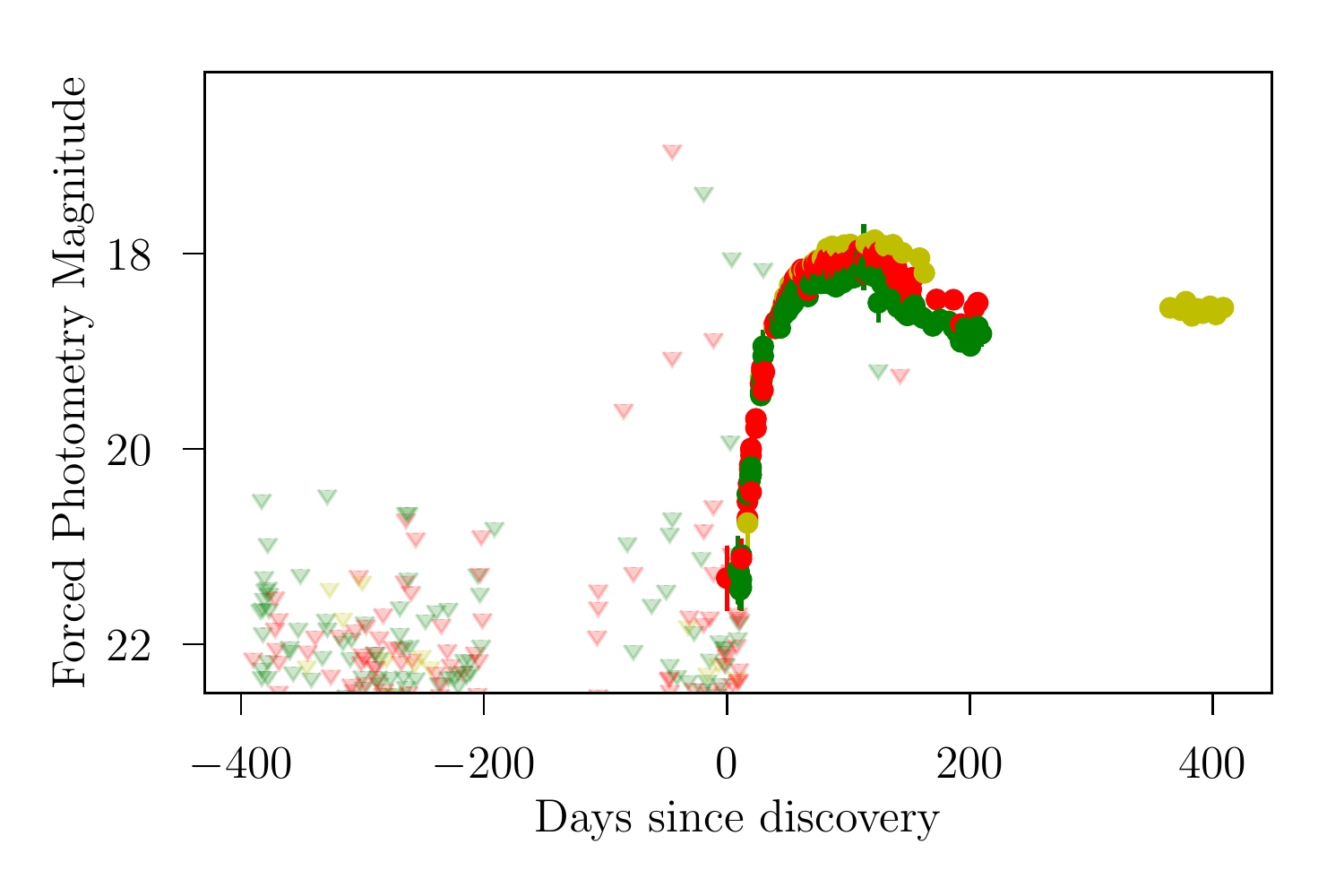}}
  
  \subfloat[\selyse]{\label{fig:ipac_lc_selyse}\includegraphics[scale=0.55]{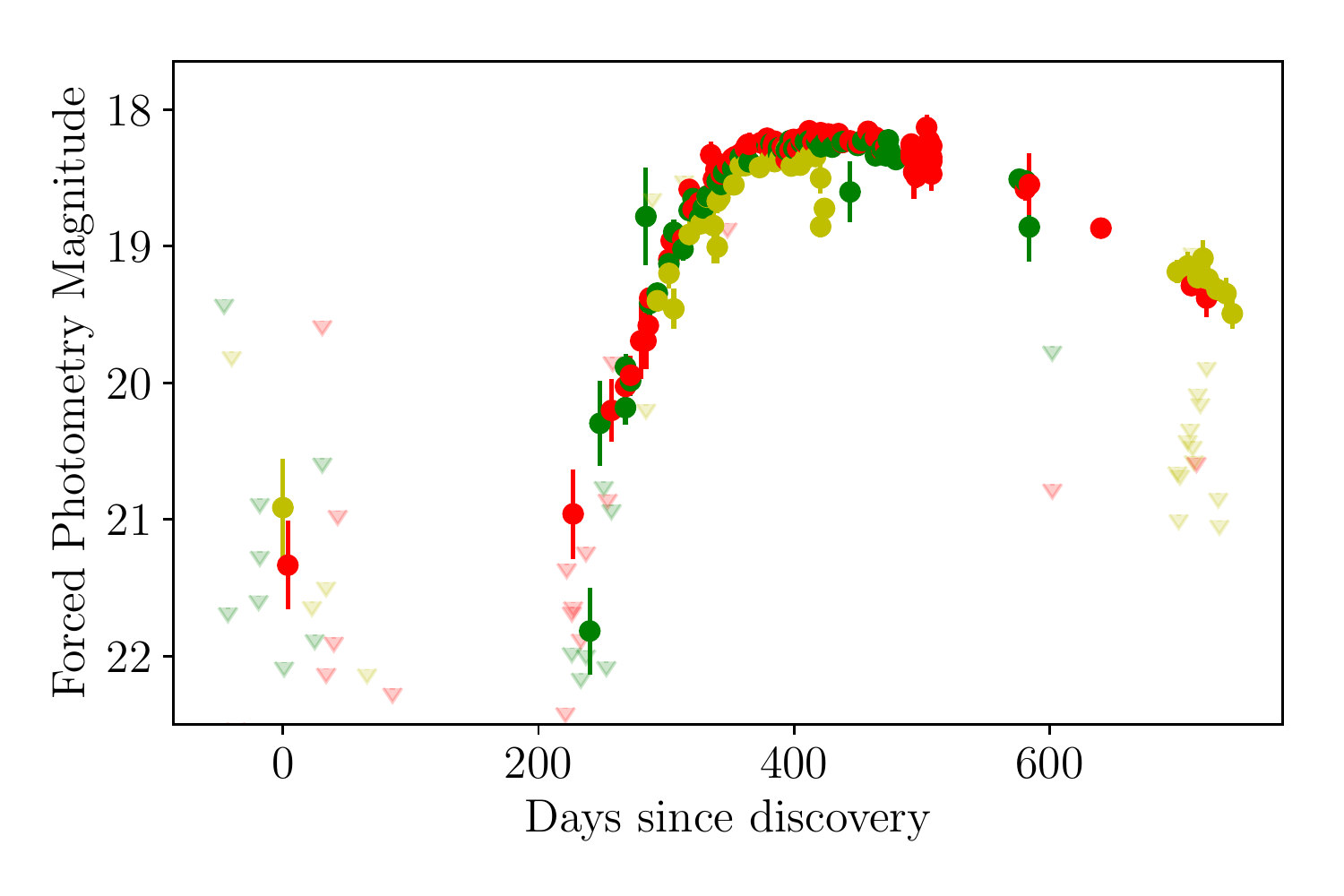}}
  \subfloat[\ciri]{\label{fig:ipac_lc_ciri}\includegraphics[scale=0.55]{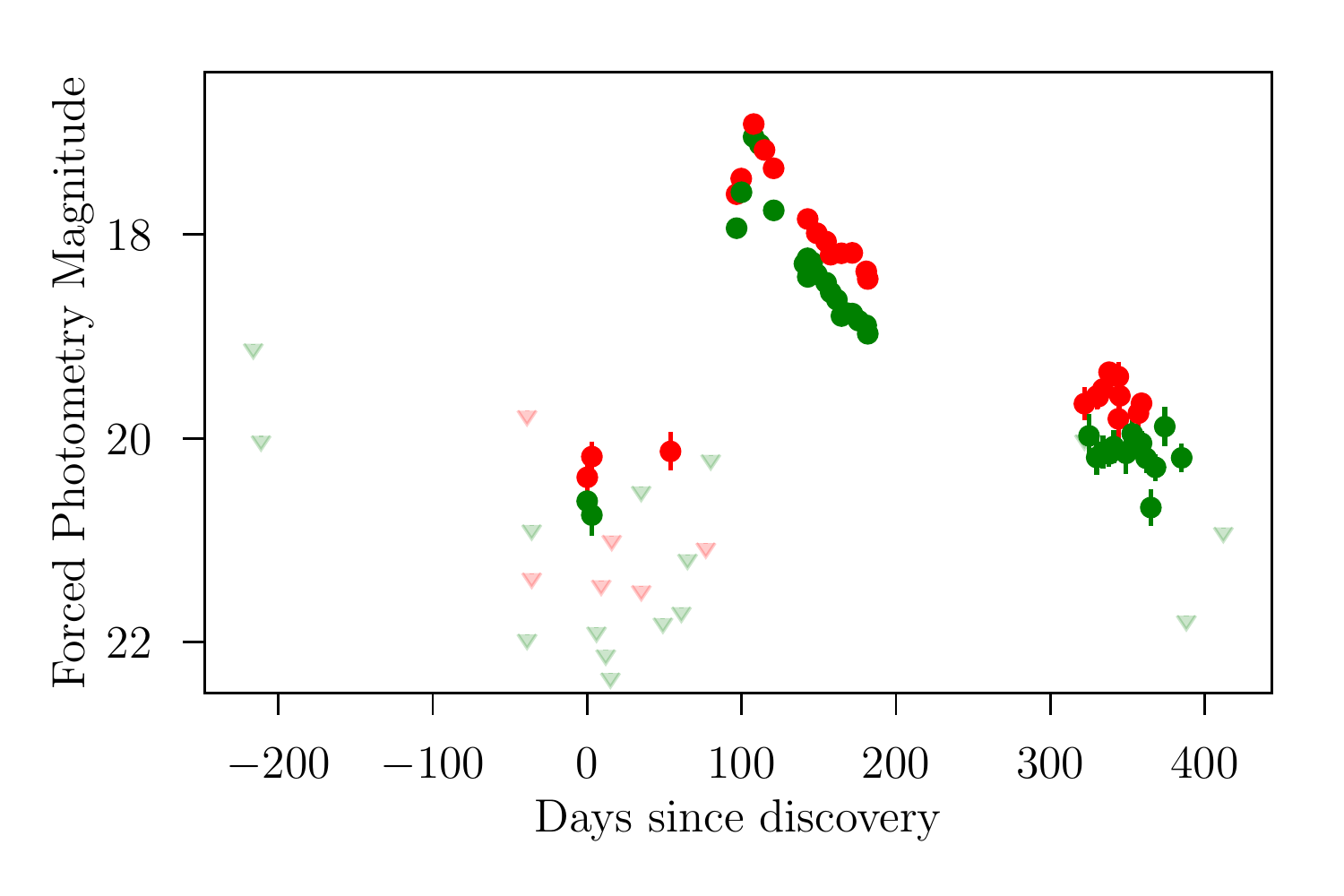}}

  \subfloat[\renly]{\label{fig:ipac_lc_renly}\includegraphics[scale=0.55]{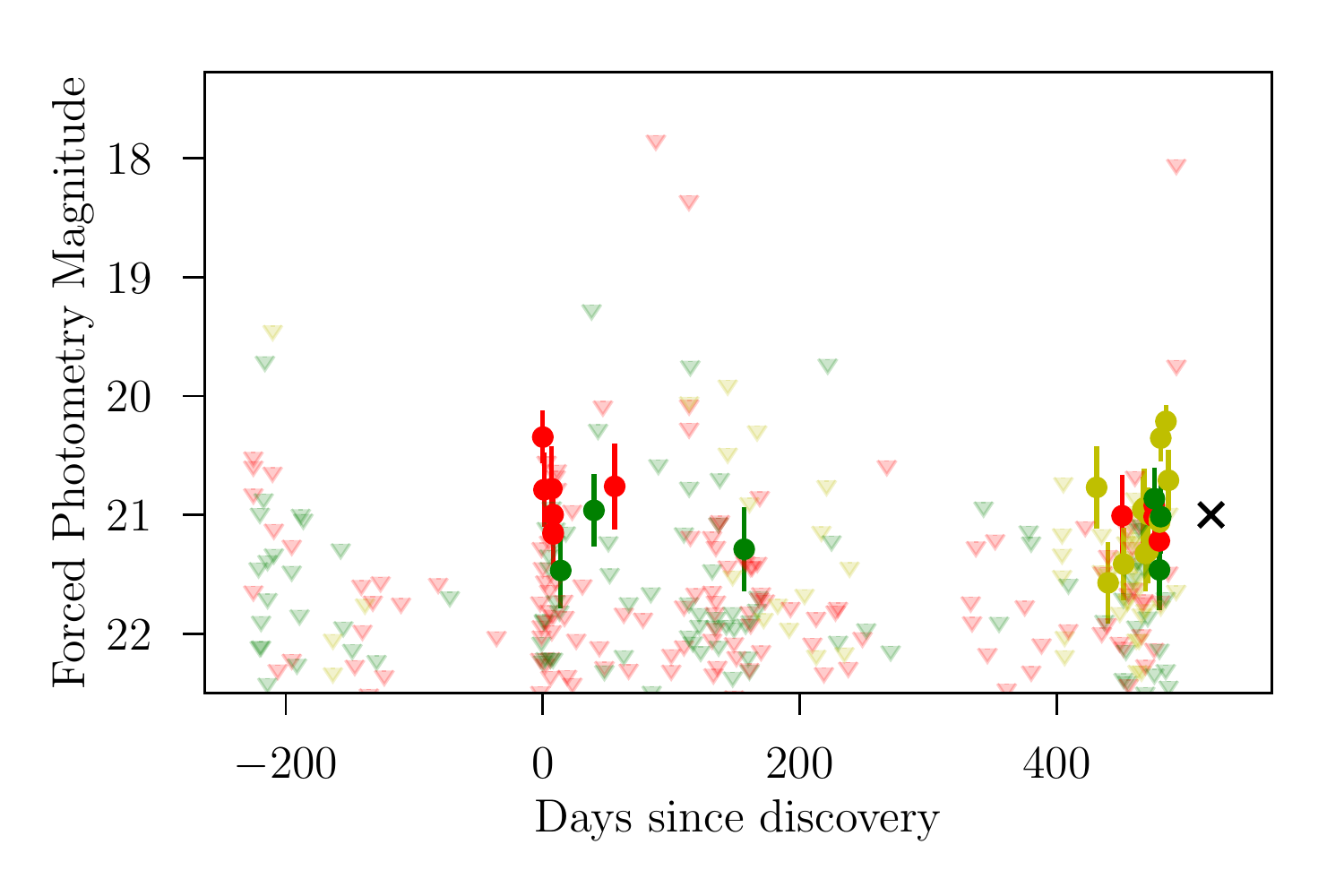}}
\caption{Forced photometry of the sample from ZTF Data Release 3. Colors correspond to $r$-, $g$-, and $i$-band 3-$\sigma$ detections, and triangles correspond to 5-$\sigma$ upper limits. An `X' marks the rise to peak in the difference imaging light curve of \renly, which was discovered in data too recent to be included in the ZTF DR3, and therefore only shows the flux level of the host galaxy. The data points in the light curves beyond 2020 will be released in the final ZTF photometry data release.}\label{fig:ipac_lc} 

\end{figure*}

\begin{figure*}[ht!]
\includegraphics[scale=0.53]{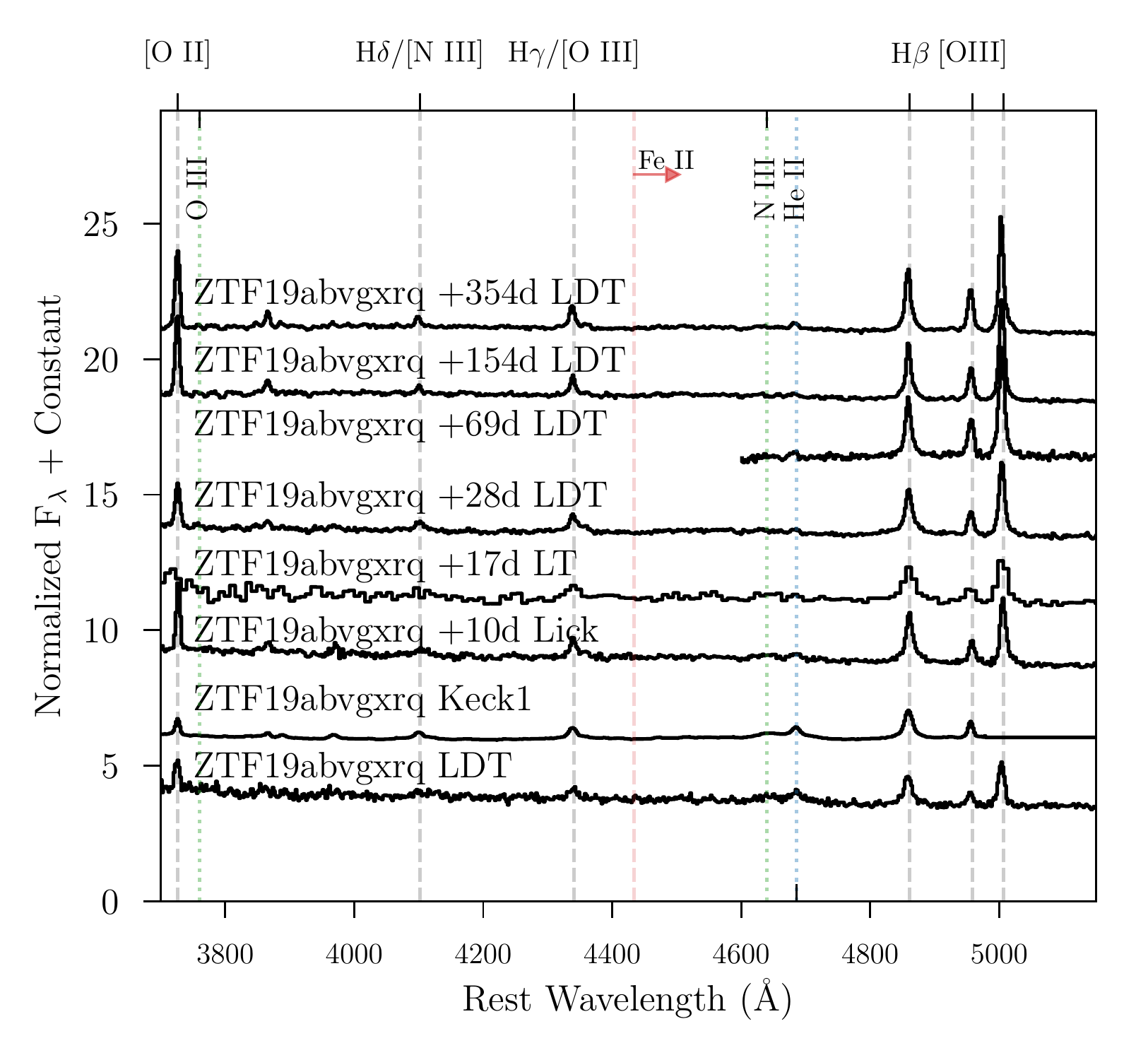}
\includegraphics[scale=0.53]{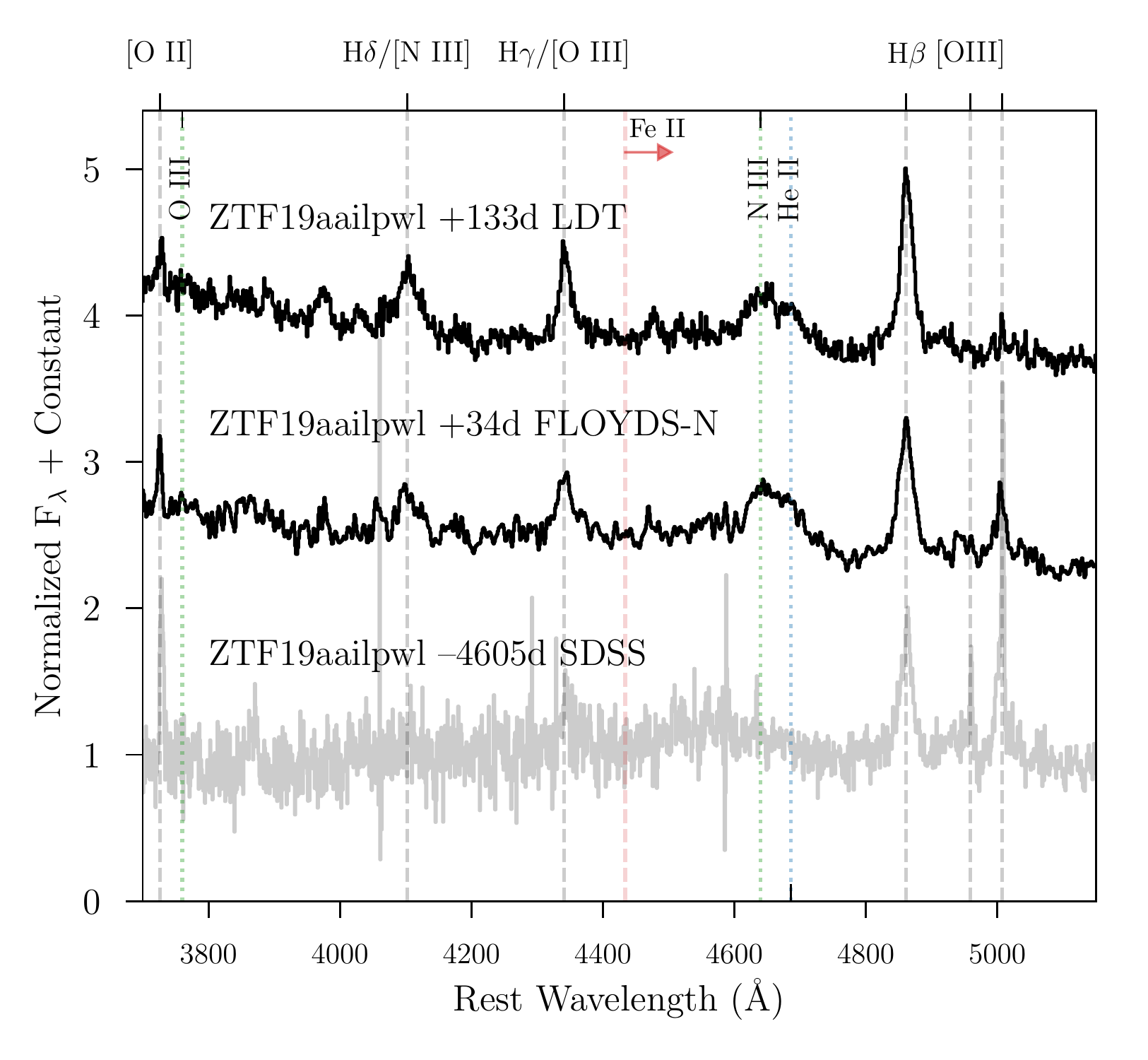}
\includegraphics[scale=0.53]{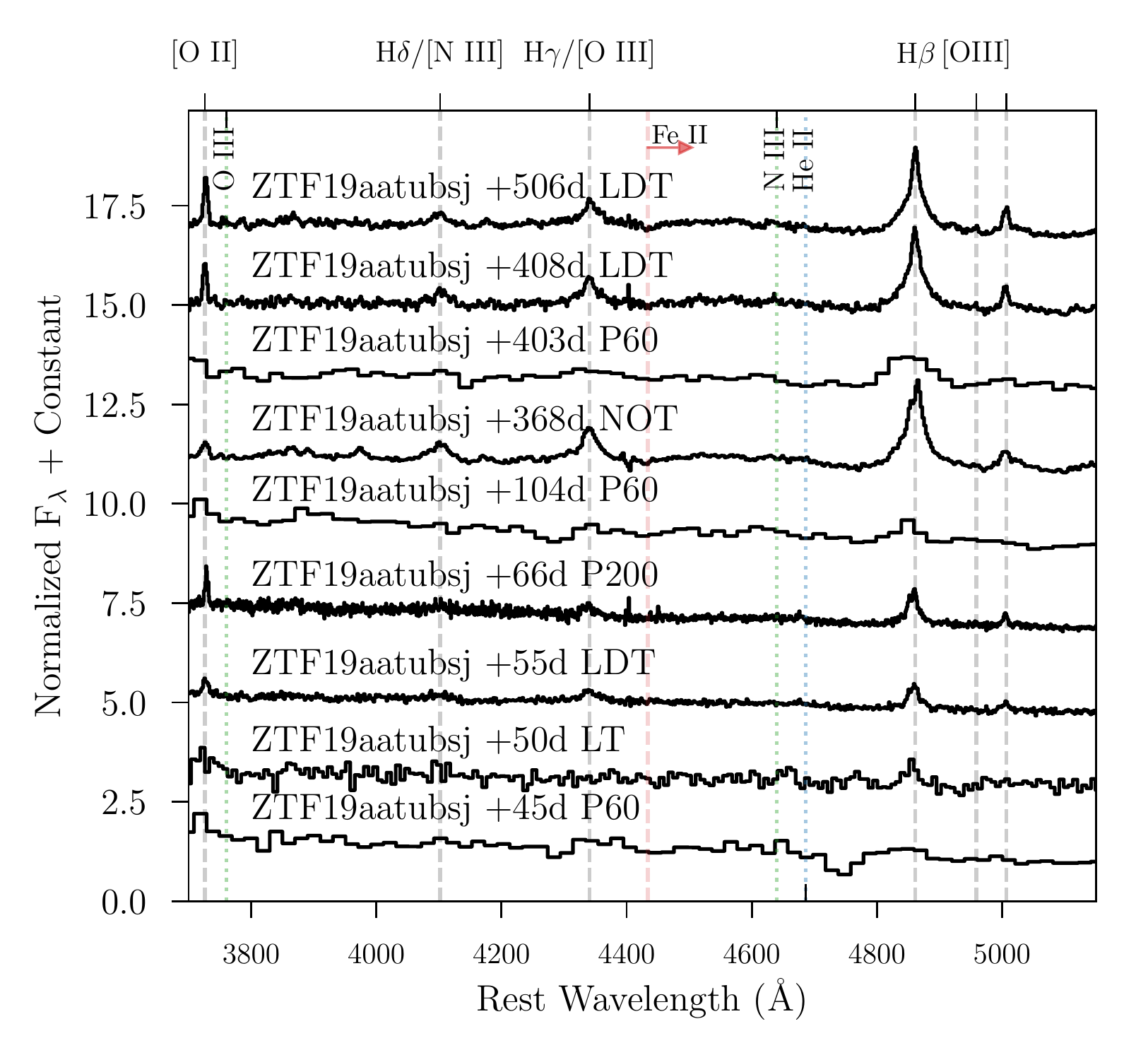}
\includegraphics[scale=0.53]{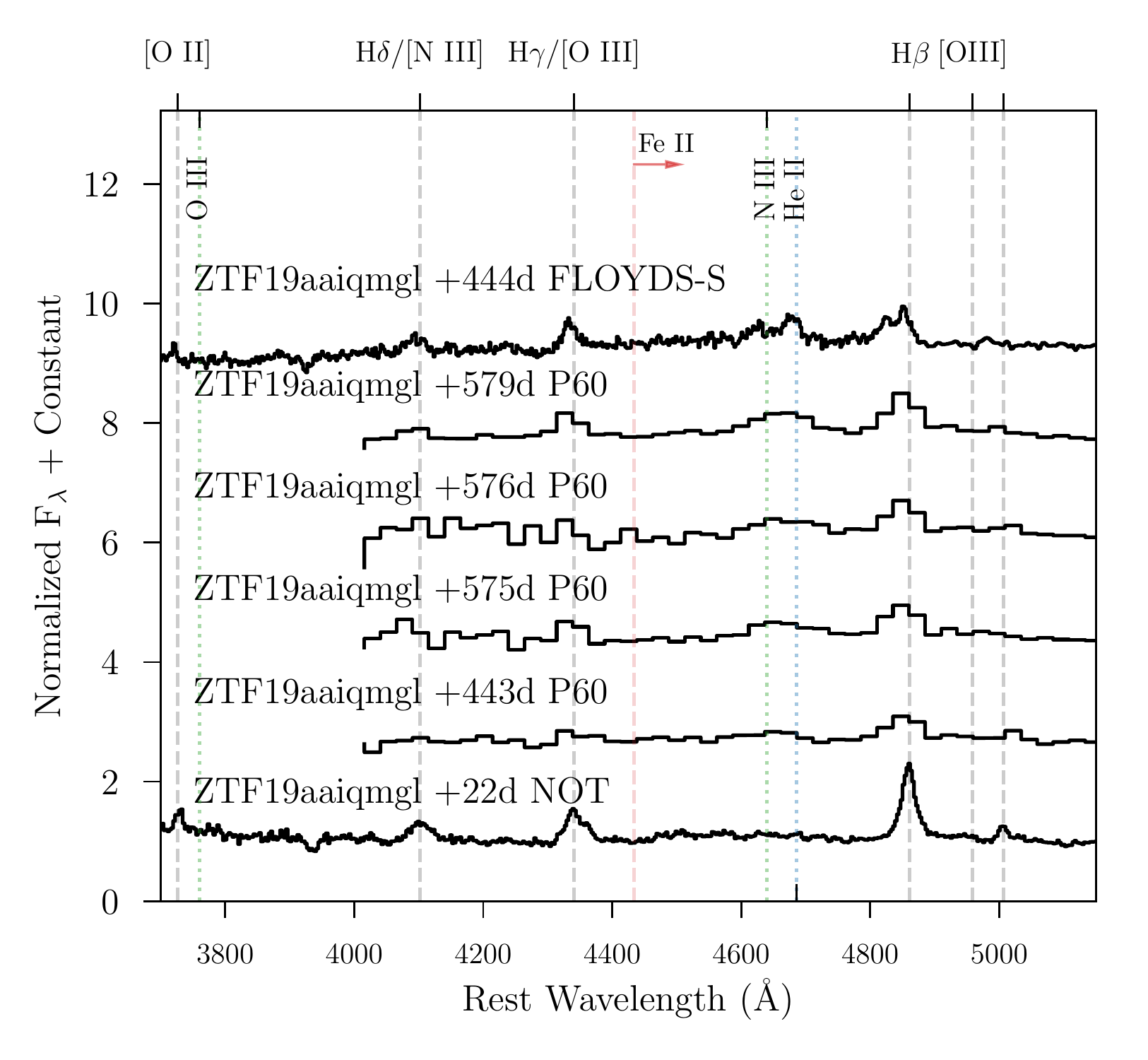} 
\includegraphics[scale=0.53]{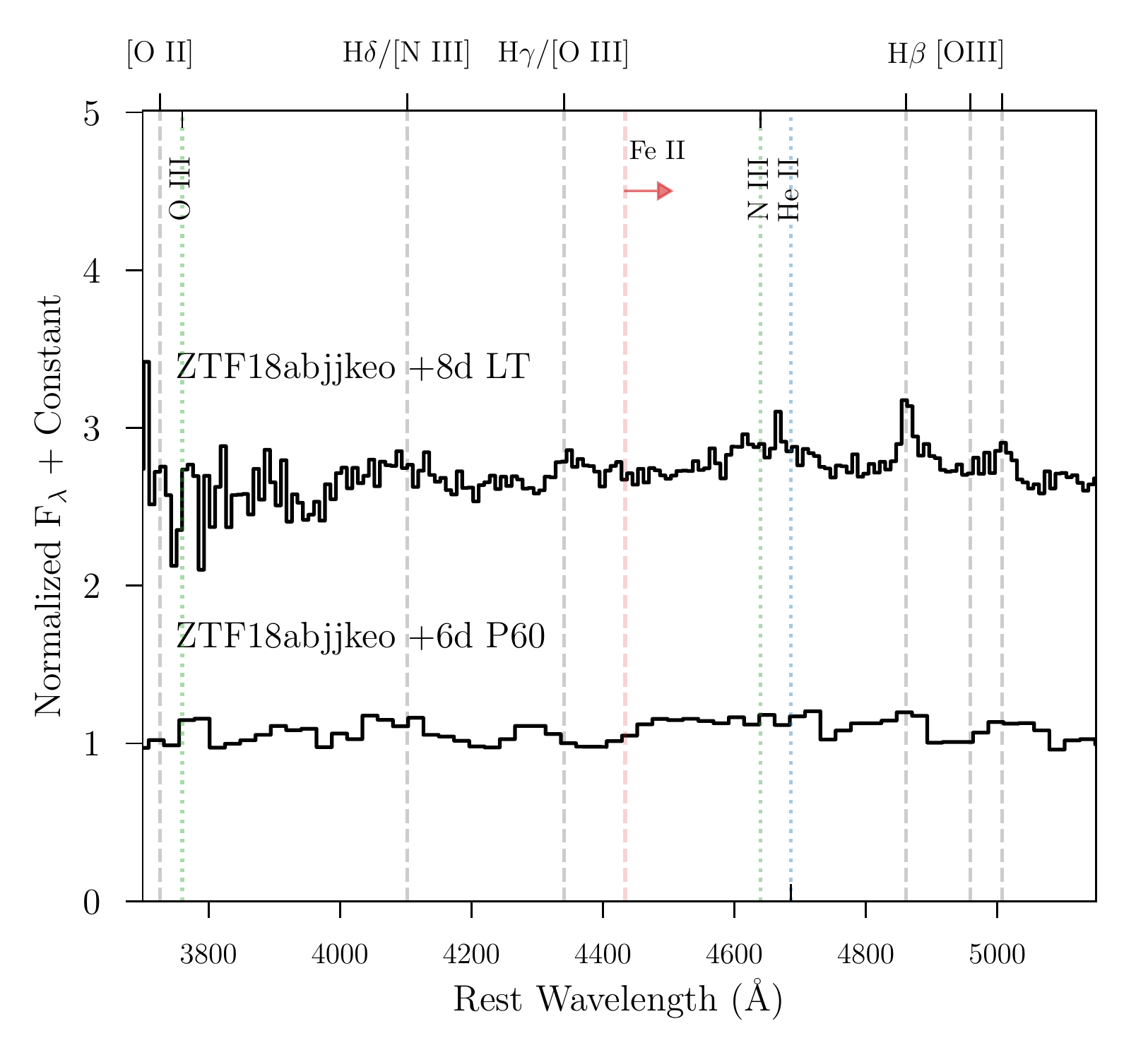}

\caption{Spectroscopic follow-up of the sample summarized in Table~\ref{tab:spc}, showing the evolution of the He~II, H$\beta$, and Fe~II line complexes. }
\label{fig:heii_spc}
\end{figure*}

The light curves in Figure~\ref{fig:ipac_lc}  are from the second IPAC data release of ZTF forced photometry. 
Figure~\ref{fig:heii_spc} shows the region of interest around He~II, H$\beta$+[O~III], and the Fe~II complex for all follow-up spectra taken of the sample.





\end{document}